\documentclass[conference]{IEEEtran} 

\usepackage{epsfig} 
\usepackage{graphicx}
\usepackage{floatrow}
\usepackage{amssymb}
\usepackage{amsfonts}
\usepackage{latexsym}
\usepackage[cmex10]{amsmath}
\usepackage[usenames]{color}
\usepackage{array}
\usepackage{cite} 
\usepackage{import}  
\usepackage[squaren]{SIunits}
\usepackage{grffile}
\usepackage{multirow}
\usepackage{booktabs} 
\usepackage{epstopdf}
\usepackage{import}
\usepackage{color}

\newcommand{\be}{\begin{equation}} 
\newcommand{\ee}{\end{equation}}
\newcommand{\ba}{\begin{array}}
	\newcommand{\ea}{\end{array}}
\newcommand{\bdm}{\begin{displaymath}}
\newcommand{\edm}{\end{displaymath}}
\newcommand{\bea}{\begin{eqnarray}}
\newcommand{\eea}{\end{eqnarray}}
\newcommand{\bean}{\begin{eqnarray*}}
	\newcommand{\eean}{\end{eqnarray*}}
\newcommand{\me}{\text{e}}

\newcommand{\mj}{\text{j}}
  
\newenvironment{proof}{\paragraph{Proof:}}{\hfill$\square$}

\def\mathlette#1#2{{\mathchoice{\mbox{#1$\displaystyle #2$}}%
                           {\mbox{#1$\textstyle #2$}}%
                           {\mbox{#1$\scriptstyle #2$}}%
                           {\mbox{#1$\scriptscriptstyle #2$}}}} 
\renewcommand{\Vec}[1]{\mathlette{\boldmath}{#1}}

\title{Real-Time Vehicular Wireless System-Level Simulation}

\author{
	\IEEEauthorblockN{Anja Dakić\textsuperscript{}, Markus Hofer\textsuperscript{}, Benjamin Rainer\textsuperscript{}, Stefan Zelenbaba\textsuperscript{}, Laura Bernadó\textsuperscript{}, Thomas Zemen\textsuperscript{}}
	\IEEEauthorblockA{
		\\\textsuperscript{}\textit{Center for Digital Safety \& Security}, \textit{Austrian Institute of Technology GmbH}, Vienna, Austria \\
		anja.dakic@ait.ac.at}
}

\begin{document}

\maketitle

\begin{abstract} Future automation and control units for advanced driver assistance systems (ADAS) will exchange sensor and kinematic data with nearby vehicles using wireless communication links to improve traffic safety. In this paper we present an accurate real-time system-level simulation for multi-vehicle communication scenarios to support the development and test of connected ADAS systems. The physical and data-link layer are abstracted and provide the frame error rate (FER) to a network simulator. The FER is strongly affected by the non-stationary doubly dispersive fading process of the vehicular radio communication channel. We use a geometry-based stochastic channel model (GSCM) to enable a simplified but still accurate representation of the non-stationary vehicular fading process. The propagation path parameters of the GSCM are used to efficiently compute the time-variant condensed radio channel parameters per stationarity region of each communication link during run-time. Five condensed radio channel parameters mainly determine the FER forming a parameter vector: path loss, root mean square delay spread, Doppler bandwidth, $K$-factor, and line-of-sight Doppler shift. We measure the FER for a pre-defined set of discrete grid points of the parameter vector using a channel emulator and a given transmitter-receiver modem pair. The FER data is stored in a table and looked up during run-time of the real-time system-level simulation. We validate our methodology using empirical measurement data from a street crossing scenarios demonstrating a close match in terms of FER between simulation and measurement.
\end{abstract}



\section{Introduction}
\label{sec:introduction}

Wireless vehicular communication systems are a key component to improve road-safety and reach zero casualties with active collision avoidance by connected advanced driver assistance systems (ADAS). For this purpose, vehicles need to exchange kinematic vehicle data and environment status information with low-latency and high reliability. Radio propagation conditions between vehicles change rapidly, since both, transmitter and receiver, are moving, which results in a non-stationary time- and frequency-selective fading process \cite{Bernado10}. Often, the direct propagation path between transmitter and receiver is blocked by buildings or other vehicles, leading to harsh wireless communication conditions. Proper testing of the low-latency real-time data exchange between multiple vehicles is a key requirement for robust operation in real road conditions.

Road tests require a lot of resources, are expensive, and are difficult to repeat since many variables cannot be controlled in a real world test. Therefore, an accurate, repeatable and resource efficient real-time system-level test methodology is needed. In a laboratory environment we shall be able to reproduce different traffic scenarios, vary simulation parameters and thus, investigate vehicular communication under realistic and repeatable wireless channel conditions.
 
In this paper\footnote{This paper is the corrected version of \cite{Dakic21} taking the Erratum \cite{Dakic21b} into account. In \cite{Dakic21} we discovered an error in the specific version of the AIT real-time channel emulator \cite{Hofer19} used to obtain the results in \cite{Dakic21}. Due to a mistake in the conversion from fixed point format 16.16 bits to 14.18 bits in the imaginary part of the (DPS sequences) coefficients, the imaginary part was divided by 4. Hence,  \cite[Figure 9-12]{Dakic21} and  \cite[Figure 16]{Dakic21} need to be updated to reflect the correct results. The result discussion for \cite[Figure 12]{Dakic21} and the signal-to-noise ratio (SNR) computation had to be adapted as well.}, we present a real-time system-level simulation method that is system independent. Hence it can be used for cellular 4G and 5G systems as well as for wireless local area network standards such as IEEE 802.11p. As special case we will show results for the IEEE 802.11p standard \cite{IEEEstandard} (using the Cohda Wireless MK5 modems). 

A numeric simulation of the IEEE 802.11p physical layer (PHY), analysing the bit error rate (BER) vs. signal-to-noise ratio (SNR) for two different channel models is presented in \cite{Mecklenbrauker11}. These channel models are a Rayleigh fading channel model with exponentially decaying power delay profile (PDP) and a geometry-based stochastic channel model (GSCM). In \cite{Bernado10} the performance of IEEE 802.11p in terms of frame error rate (FER) vs. SNR is presented for different channel estimation and equalization techniques. Another evaluation of the IEEE 802.11p PHY performance is presented in \cite{Anwar19} focusing on a comparison of different vehicular communication technologies in terms of packet error rate with respect to latency, reliability and data rates. In \cite{Bernado10} a tapped-delay line (TDL) channel model for an urban non line-of-sight (NLOS) street crossing scenario is utilized.  

In \cite{Vlastaras15}, a real-time wireless channel emulator is proposed for testing an IEEE 802.11p transceiver for scenarios with large root mean square (RMS) Doppler- and RMS delay spread. The authors in \cite{Blazek19} propose to represent a large vehicular communication network with an interference channel and a single TDL channel emulator. In \cite{Bazzi20}, a hardware-in-the-loop (HiL) simulation platform is proposed, where a traffic simulator, a signal generator and a TDL channel emulator are combined. In \cite{Anwar18}, physical layer abstraction techniques for IEEE 802.11p and LTE vehicle-to-vehicle (V2V) based on a TDL channel model for system-level simulation are investigated.  

All above mentioned prior work suffer from simplified propagation models that do not fully take the non-stationary properties of the vehicular communication channel into account. It is well known \cite{Molisch10} that the FER is strongly affected by the non-stationary properties of the vehicular communication channel. In this work, we pursuit the goal of accurate real-time system-level simulation, hence we want to relate the PHY layer properties directly to the FER that can be supplied to a network simulator.

In order to capture the non-stationary fading process of vehicular scenarios properly, GSCMs \cite{Karedal09} offer a good trade off between complexity and accuracy. Due to the non-stationarity of the fading process, the central condensed radio channel parameters are time-varying and can be summarized as path loss, RMS delay spread, RMS Doppler spread, Rician $K$-factor (in the text for simplicity referred to as $K$-factor) and Doppler shift of the line-of-sight (LOS) component. The condensed radio channel parameters will be constant only for a stationarity region \cite{Bernado14} with an extend of several wave lengths.

In this paper, we investigate the hypothesis that the time-variant condensed channel parameters of a non-stationary fading channel can be locally approximated for each stationarity region by the statistics of those from a wireless channel generated by a stochastic channel model with an exponentially decaying PDP and Clark's Doppler power spectral density (DSD) \cite{Clarke68} with an additional LOS component. Thus, enabling us to obtain the FER using HiL measurements for different condensed radio channel parameters, which can then be used as lookup table during run-time of the system-level simulation.


\subsection*{Scientific Contributions of the Paper}
\begin{itemize}
    \item We propose a HiL test methodology for obtaining the FER of a specific modem hardware. The channel emulator used in the HiL setup implements a stochastic channel model where we can specify five key condensed radio channel parameters of the generated channel impulse responses: Path loss, RMS delay spread, Doppler bandwidth, $K$-factor, and LOS Doppler shift. 
	
	\item We measure the FER for a pre-defined set of discrete grid points of the condensed radio channel parameter vector to fill a lookup table indexed by those values. This FER lookup table is later used for system-level simulation. The FER measurements are performed before run-time once for a given modem pair. 
	
	\item We propose a simple and fast approach to calculate the condensed radio channel parameters for each stationarity region directly from the propagation path parameters of the GSCM. Hence, during run-time of the system-level simulation, the resulting channel impulse responses do not need to to be computed explicitly leading to a strong complexity reduction.
	
	\item We use the condensed radio channel parameters for indexing the FER lookup table to obtain the FER of a specific communication link, thus reducing the computational complexity of a multi-node system-level simulator achieving real-time operation for the first time.
\end{itemize}

\subsection{Organization of the Paper} 
We present the methodology of our system-level simulation in Sec.~\ref{sec:methodology}. In Sec.~\ref{sec:LUTgenerator}, we describe the HiL test setup and explain the process for creating the FER lookup table. Our system-level simulation and the computational complexity reduction approach is shown in Sec.~\ref{sec:SLS}. We present the results of the HiL measurements, validate our model and discuss results in Sec. \ref{sec:resanddiscusion}. In Sec. \ref{sec:conlcusion} conclusions are drawn.

\subsection{Notation}
We denote a scalar by $a$, the absolute value of $a$ by $\left|a\right|$ and its complex conjugate by $a^*$. A vector is denoted by $\Vec{a}$. Furthermore, $a(\cdot)$ is used for continuous parameters, while $a[\cdot]$ is used for discrete parameters. The expected value of a random value $X$ is given by $\mathbb{E}[X]$. We denote the set of all real numbers by $\mathbb{R}$ and of all complex numbers by $\mathbb{C}$, respectively.

\section{Methodology}
\label{sec:methodology}

\begin{figure*}[ht] 
	\begin{center}
    \includegraphics[width=1\textwidth]{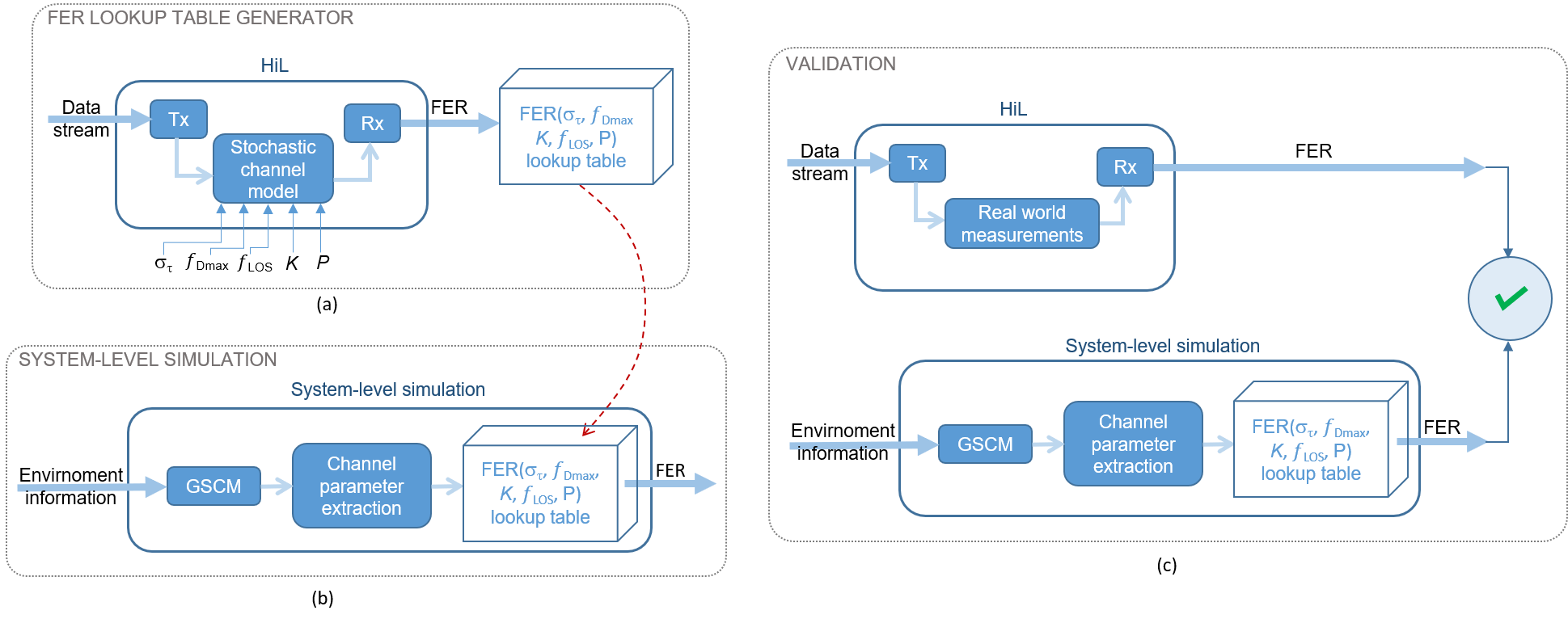}
    	\end{center}
	\caption{Methodology of our approach and its validation. (a) FER lookup table generator: A hardware-in-the-loop (HiL) setup for generating the FER lookup table. The channel impulse response from a stochastic channel model with the condensed channel parameter vector $\Vec{\Psi}$ (RMS delay spread $\sigma_\tau$, Doppler bandwidth $f_\text{Dmax}$, LOS Doppler shift $f_\text{LOS}$, $K$-factor $K$, and received power $P$) is emulated. (b) System-level simulation: GSCM module for simulating the environment and computing the propagation paths; channel parameter extraction module for estimating the condensed channel parameter vector $\Vec{\Psi}'$ and FER lookup table for obtaining the FER during run-time. (c) Validation: The FER obtained by emulating the channel impulse response from real world measurements is compared with the FER obtained by the system-level simulator.}
\label{fig:process}
    \end{figure*}

For real-time system-level simulation, we follow a two-step approach depicted in Figure~\ref{fig:process}~(a) and~(b):
\begin{enumerate}
    \item \textit{FER Lookup Table Generation:} First, we create a FER lookup table using a HiL setup, explained in detail in Sec.~\ref{sec:LUTgenerator} and depicted in Figure~\ref{fig:process}~(a). We choose a stochastic channel model with an exponentially decaying PDP, considering Rician fading in the first delay tap and Rayleigh fading in all other taps, to be emulated by our AIT channel emulator \cite{Hofer19}. The condensed radio channel parameters of this model can be kept constant for arbitrary long periods enabling accurate HiL FER measurements of any radio transmitter-receiver pair. Specifically the condensed channel parameter vector
\begin{equation}
\Vec{\Psi}= 
\left[ P, \sigma_\tau, f_\text{Dmax},  K, f_\text{LOS} \right]^\text{T}
\end{equation}
contains the received power $P$, RMS delay spread $\sigma_\tau$, Doppler bandwidth $f_\text{Dmax}$, $K$-factor of the first delay tap $K$, and the LOS Doppler shift $f_\text{LOS}$, hence creating channel impulse responses under different LOS and non LOS (NLOS) conditions. The overall path loss $\rho$ and the transmit power $P_\text{Tx}$ determines the received power $P$
\begin{equation}
P = P_\text{Tx}+\rho.
\label{eq:Rxpower}
\end{equation}

 We assume that the FER of this channel model with parameter vector $\Vec{\Psi}$ is a sufficiently good approximation to the FER of a non-stationary fading radio channel with a similar parameter vector $\Vec{\Psi}'$ within one stationarity region \cite{Molisch10}. All FER values on a grid $\mathcal{I}= I_P \times I_{\sigma_\tau} \times I_{f_\text{Dmax}} \times I_K \times I_{f_\text{LOS}}$ are measured and stored before run-time. Here, $I_x$ denotes a set with a suitable discretization of the parameter range for a certain scenario, see Table \ref{table2}.
 
\item \textit{System-Level Simulation:}We load the obtained FER lookup table in a system-level simulator that consists of three modules: a GSCM for generating propagation paths, a condensed channel parameter estimation module, and the FER lookup table, see Figure~\ref{fig:process}~(b). For simulating the time and frequency selective wireless channel we use a GSCM. However, obtaining the FER of a transmission for a specific wireless communication system using a GSCM is computationally expensive and cannot be computed in real-time for multi-node and complex non-stationary scenarios (e.g., street canyons, urban areas). Thus, we aim to efficiently approximate the condensed radio channel parameters from the propagation paths generated by a GSCM within one stationarity region, described in detail in Sec.~\ref{subsec:Computationparameters}. During run-time of the system-level simulator, the parameter vector $\Vec{\Psi}'$ is computed for each stationarity region of the non-stationary fading channel and the closest matching entry in the lookup table supplies the FER for the network protocol simulator.

In this paper, we show a detailed description of our methodology for a single communication link. To extend this to multiple communication links, we have to obtain the parameter vector $\Vec{\Psi}'$ and use it as the index into the FER lookup table for each stationarity region of all present links in parallel. With our implementation on an Intel(R) Xeon(R) Gold 6150 CPU it can be done for $30$ simultaneously simulated communication links without any limitation.

\end{enumerate}

\subsection{Channel model generalities}
\label{subsec:channelmodel}
The impulse response of the physical wireless channel can be written as \cite{Hlawatsch11}
\begin{equation}
h_\text{ph}(t,\tau) = \sum_{l=1}^L\eta_l(t)\delta_\text{D}(\tau-\tau_l(t)),
\label{phCIR}
\end{equation}
where $t$ represents time and $\tau$ delay, respectively. Each path $l$, with $l\in\left\lbrace 1,...L\right\rbrace $ and $L$ being the total number of multi-path components (MPCs), has a time-variant delay $\tau_l(t) \in \mathbb{R}$ and a complex time-variant weighting coefficient $\eta_l(t)$, which is defined as $\eta_l(t)= \eta'_l(t)\me^{j2\pi \phi_l} \in \mathbb{C}$, with $\eta'_l(t) \in \mathbb{R}$ being the amplitude and $\phi_l$ the starting (random) phase of the $l$-th path \cite{Hofer19}. 


Since $h_\text{ph}(t,\tau)$ is not band-limited, we employ band-limiting filters at the transmitter and receiver side with impulse responses $h_\text{Tx}(\tau)$ and $h_\text{Rx}(\tau)$, respectively. The final channel impulse response is obtained by
\begin{equation}
h(t,\tau)=h_\text{Tx}(\tau)*h_\text{ph}(t,\tau)*h_\text{Rx}(\tau),
\label{totalCIR}
\end{equation}
where $*$ denotes the convolution operator. For both filters we consider a root raised cosine filter which, in a cascade, result in a raised cosine (RC) filter with impulse response \cite{Mosa07}
\begin{equation}
h_\text{RC}(\tau) = \frac{\sin(\pi \tau/T_\text{c})}{\pi  \tau/T_\text{c}} \frac{\cos(\beta \pi \tau/T_\text{c})}{1-(2\beta \tau/T_\text{c})^2},
\label{rcfilter}
\end{equation}
where $T_\text{c}$ represents the sampling in delay and $\beta$ the roll-off factor which can be any value within $[0,1]$. 
After the convolution of the physical channel and the RC filter we get 
\begin{equation}
h(t,\tau) = \sum_{l=1}^{L} \eta_l(t)h_\text{RC}(\tau-\tau_l(t)).
\label{afterconv}
\end{equation}

Considering a maximum one-sided Doppler bandwidth $B_\text{D}$, we sample our channel impulse response in time with a repetition rate of $T_\text{s}<1/(2B_\text{D})$. This results in $M = T_{\text{stat}}/T_{\text{s}}$ samples within a stationarity region. We define a stationary time $T_{\text{stat}}$ within which we assume that the wide sense stationarity assumption of the fading process holds \cite{Matz05}. In delay we sample with $T_\text{c}=1/B$, where $B$ denotes the communication bandwidth.
Sampling \eqref{afterconv}, we obtain the discrete time channel impulse response
\begin{equation}
h[m',n] = \sum_{l=1}^{L} \eta_l[m'] h_\text{RC}(nT_\text{c} - \tau_l[m']),
\label{discereteCIR}
\end{equation}
where 
\begin{equation}
    m'=rM+m
\end{equation}
with $r\in\left\lbrace 0,\ldots, R-1\right\rbrace$ denoting the stationarity region index,  $m\in\left\lbrace 0,\ldots, M-1\right\rbrace$ denoting the time index within one stationarity region, $M$ being the total number of samples within a stationarity region,  and $n\in\left\lbrace 0,\ldots, N-1\right\rbrace$ denoting the time delay index.

For a single stationarity region $r$ with duration $T_{\text{stat}}$ we assume a negligible change in the amplitude of the propagation paths, 
\begin{equation}
|\eta_l[rM+m]| \approx |\eta_{l,r}|
\label{eq:ConstWeight}
\end{equation} 
a constant relative velocity $v_{l,r}$, and a constant relative angle between transmitter and receiver. These assumptions lead to a constant Doppler shift $f_{l,r}$ during $T_{\text{stat}}$. Further we can write
\begin{equation}
    \tau_{l}[m'] = \tau_{l,r} - \frac{v_{l,r}}{c_0} m T_{\text{s}}
    \label{eq:ConstVel}
\end{equation}
where $\tau_{l,r}=\tau_l[rM]$ and $c_0$ denotes the speed of light. Finally, we can write \cite{Hofer19}
\begin{equation}
h[m',n] \approx \sum_{l=1}^{L} |\eta_{l,r}| \me^{j2 \pi (\phi_{l,r} - \nu_{l,r} m)} h_\text{RC}(nT_\text{c} - \tau_{l,r}).
\label{discereteCIR2}
\end{equation}
where $|\eta_{l,r}|$ represent the constant channel weight for stationarity region $r$, $\phi_{l,r}$ the starting phase, $\nu_{l,r} = v_{l,r} f_\text{C} T_\text{s}/c_0$ the normalized Doppler shift of path $l$, and the Doppler shift $f_{l,r}= \nu_{l,r}/T_\text{s}$. 

\section{FER lookup table generation}
\label{sec:LUTgenerator}

\subsection{Stochastic Channel Model for Frame Error Rate Estimation}
\label{subsec:TDLModel}
A stochastic channel model is used for the estimation of the FER lookup table entries. We consider the fading process to be wide sense stationary (with uncorrelated scattering) for a given period of time. This allows us to derive closed forms for the statistics of the fading process which are determined by the model parameters. We consider a specific number of delay taps which we model with an exponentially decaying PDP 
\begin{equation}
|\alpha_{\text{exp}}[n]|^2=
\begin{cases}
\me^{-n\Delta t/\tau_0}, &\text{$1\leq n \leq N_\text{taps}$}\\
0, &\text{$n < 1\, $}\, 
\end{cases}.
\label{pdpdiscrete}
\end{equation}
where $\Delta t=T_\text{c}$ denotes the tap delay spacing, $|\alpha_\text{exp}[n]|^2$ the power of tap $n$ and $\tau_0$ the delay parameter, respectively. The delay of the tap $n$ is $\tau_n=n\Delta t$, for $n \in \left\lbrace 1,.., N_\text{taps}\right\rbrace $, with $ N_\text{taps}=\lceil \frac{\tau_\text{max}}{T_\text{c}}\rceil$. In Appendix \ref{app:RMSDs} the calculation of the RMS delay spread for an exponential PDP with a given maximum delay $\tau_\text{max}$ is shown.

To obtain a normalized PDP we use
\begin{equation}
|\alpha_\text{exp}^\prime[n]|=\frac{|\alpha_\text{exp}[n]|}{\sqrt{P_\text{total}} }   ,
\label{eq:normalization2}
\end{equation}
where $P_\text{total}$ is the sum power of all taps
\begin{equation}
\label{eq:totalpower2}
P_\text{total}=\sum_{n=1}^{N_\text{taps}}|\alpha_\text{exp}[n]|^2.
\end{equation}

\begin{figure}[t!]    
	\centering
	\includegraphics[width=0.50\columnwidth]{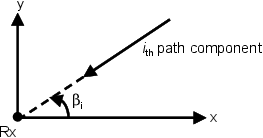}
	\caption{Angle of arrival $\beta_i$ of the $i_\text{th}$ path/plane wave component.}
	\label{fig:AoA}
\end{figure}

The channel is described by $ N_{\text{taps}}$ delay taps, where each of the taps consists of $L'$ MPCs exhibiting the same delay. Every MPC arrives at the receive antenna from a different angle $\beta_{i}$ with a specific attenuation and delay, as is shown in Figure~\ref{fig:AoA}. We assume a random uniformly distributed starting phase ($\phi_i \underset{\text{i.i.d}}{\sim} U(0,2\pi)$) for each MPCs. The channel impulse response 
\begin{align}
&    h[m, n] = \sqrt{\frac{K}{K+1}} \me^{-\mj 2 \pi f_{\text{LOS}} m T_s + \mj\phi_{\text{LOS}}} \delta_\text{K}(n-1)  \nonumber \\
& +  \sqrt{\frac{1}{(L'-1)(K+1)}} \sum_{i=1}^{L'-1} \me^{-\mj 2 \pi \rho_{1,i} m T_s + \mj\phi_{1,i}} \delta_\text{K}(n-1)  \nonumber \\ 
&+ \sqrt{\frac{1}{L'}} \sum_{w=2}^{N_{\text{taps}}}  \sum_{i=1}^{L'}  \me^{-\mj 2 \pi \rho_{w,i} m T_\text{s} + \mj\phi_{w,i}}  \delta_\text{K}(n-w)\,, 
\label{eq:generatedChannlNormalized}
\end{align}
where $\rho_i = f_{\text{Dmax}} \cos(\beta_i)$ and $\delta_\text{K}$ denotes the Kronecker delta. The angle of arrival is denoted by $\beta_i \underset{\text{i.i.d}}{\sim}U(-\pi, \pi)$ for a double sided, by $\beta_i \underset{\text{i.i.d}}{\sim}U(-\pi/2, \pi/2)$ for a right sided and by $\beta_i \underset{\text{i.i.d}}{\sim}U(\pi/2, 3\pi/2)$ for a left sided Doppler spectrum, respectively.

In case of a NLOS scenario $K=0$ and we model all taps as Rayleigh fading adhering to Clarke’s spectrum \cite{Clarke68}. The Doppler bandwidth of the Clarke's Doppler spectrum is $B_\text{D} = f_\text{Dmax}$ where $f_\text{Dmax}$ denotes the maximum one sided Doppler bandwidth. The RMS Doppler spread $\sigma_{\nu}$ for a Clarke's model can be obtained from the Doppler bandwidth and the following relation \cite{Clarke68}
\begin{equation}
\sigma_{\nu} = \frac{f_\text{Dmax}}{\sqrt{2}}.
\label{dopplerspread}
\end{equation}

In the case of a LOS scenario, we model only the first tap as Rician with   $K>0$. We can see in \cite{Bernado15} that only the first tap follows a Rician distribution, while the second one changes its distribution from Rician to Rayleigh, and the fading process for later taps exhibits a Rayleigh distribution. 

We denote the deterministic Doppler shift of the LOS by $f_{\text{LOS}} \in [-f_\text{Dmax}, f_\text{Dmax}]$. In the presence of such a deterministic component \eqref{dopplerspread} is no longer valid. It is also evident that with higher $K$-factors the LOS component is more dominant and, thus, the RMS Doppler spread changes depending on the $K$-factor and the Doppler shift of the LOS component. We further notice that the shape of the PDP has also an impact on the RMS Doppler spread. In order to validate these intuitive dependencies we derive the (conditional) Doppler spectral probability density with respect to $f_{\text{LOS}}$, $f_\text{Dmax}$, $K$-factor, and the employed PDP profile. We derive the Doppler spectral density for our channel model by proceeding like in \cite[Appendix C]{Clarke68}, for details see Appendix \ref{app:DSD:measChannel}. The two sided (bathtub shaped) Doppler spectral probability density $p_{\text{full}}$ is given by
\begin{align}
&    p_{\text{full}}(f| f_{\text{LOS}}, f_\text{Dmax}, K, \text{PDP}) = \nonumber \\
&    |\alpha[1]|^2 \left( \frac{K}{1+K}\delta_D(f_{\text{LOS}} - f) + \frac{1}{1+K}\frac{1}{\pi (f_\text{Dmax}^2 - f^2)^{\frac{1}{2}}} \right)  \nonumber \\
&+\frac{1}{\pi (f_\text{Dmax}^2 - f^2)^{\frac{1}{2}}}\sum_{i=2}^{N_\text{taps}} |\alpha[i]|^2,
\label{eq:channelModelDopplerSpectralDensity}
\end{align}
where 
\begin{equation}
    \sum_{i=1}^{N_\text{taps}} |\alpha[i]|^2 = 1
\end{equation}
i.e,. $|\alpha[i]| = |\alpha_{\text{exp}}^\prime[i]|$ and $f \in [-f_\text{Dmax}, f_\text{Dmax}]$.

From \eqref{eq:channelModelDopplerSpectralDensity} we obtain the variance by applying $\mathbb{E}[(X - \mathbb{E}[X])^2] = \mathbb{E}[X^2] - \mathbb{E}[X]^2$ resulting in
\begin{align}
    \sigma_{\nu}^2 &= |\alpha[1]|^2 \left(\frac{2f_{\text{LOS}}^2 K+f_\text{Dmax}^2}{2 (1+K)} - \frac{|\alpha[1]|^2 f_{\text{LOS}}^2 K^2}{(1+K)^2} \right)  \nonumber \\
& + \frac{f_\text{Dmax}^2}{2} \sum_{i=2}^{N_\text{taps}} |\alpha[i]|^2.
\label{eq:analyitcalDopplerSpread}
\end{align}

\subsection{Numerical RMS Doppler spread evaluation}

 Considering a carrier frequency $f_\text{C} = 5.9\,\text{GHz}$ and a communication bandwidth $B = 10\,\text{MHz}$  we evaluate the RMS Doppler spread by applying specific numerical values of the RMS delay spread, the LOS Doppler shift and the $K$-factor.

\begin{figure}[t!]    
	\centering
	\includegraphics[width=0.95\columnwidth]{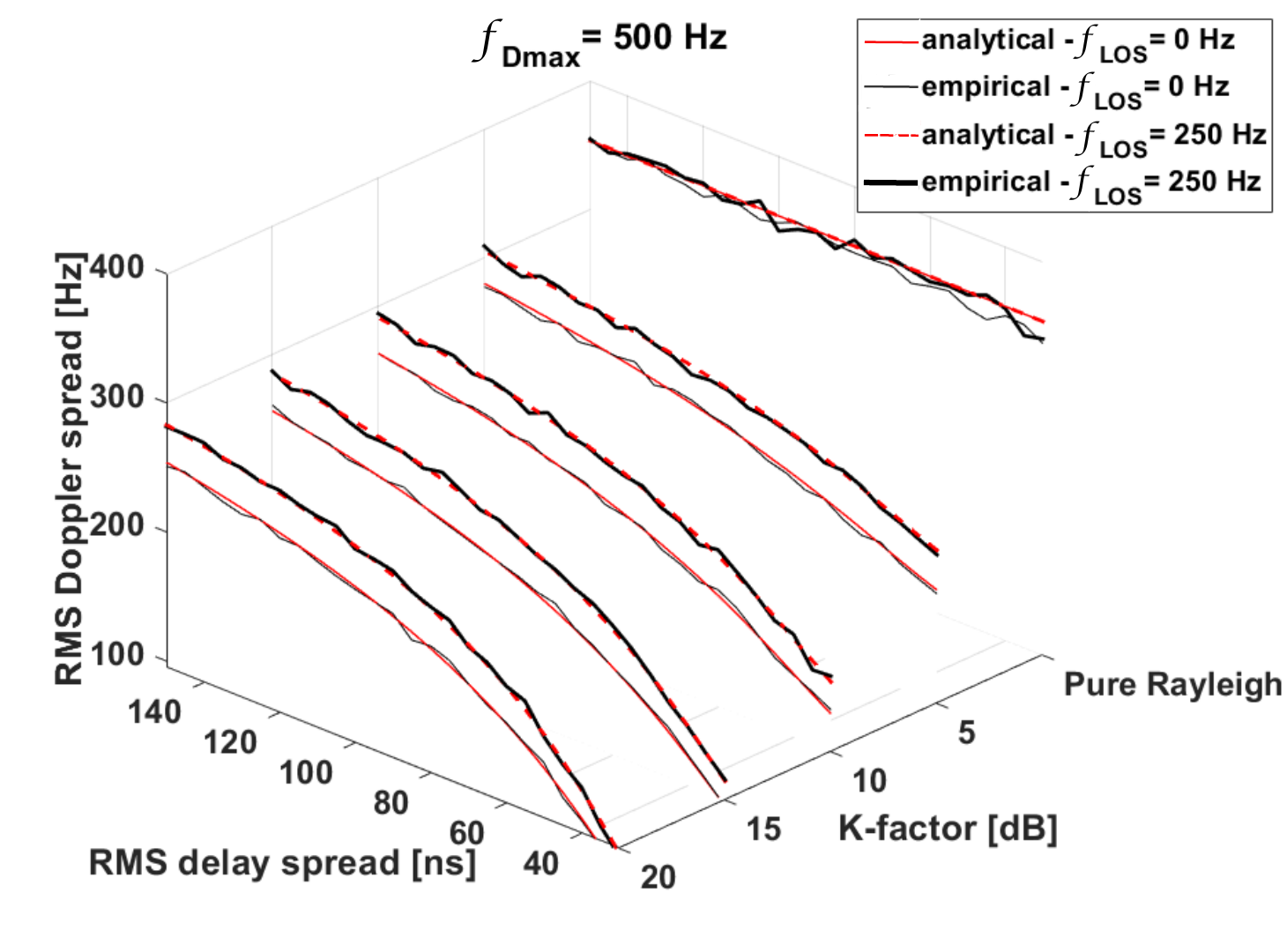}
	\caption{Analytical RMS Doppler spread compared to the empirical RMS Doppler spread obtained by averaging 20 simulation runs using an exponential PDP with delay parameter $\tau_0$ from $34\,\text{ns}$ to $150\,\text{ns}$, $8$ delay taps, $f_\text{Dmax}=500\,\text{Hz}$, $f_{\text{LOS}} \in \{0,250\}\,\text{Hz}$. The band limited channel impulse responses of the simulations is calculated by \eqref{afterconv} using a roll off of $0.9$ and a system bandwidth of $B'=10\,\text{MHz}$.}
	\label{fig:DopplerSpreadAnalyticalvsEmpirical}
\end{figure}

Figure~\ref{fig:DopplerSpreadAnalyticalvsEmpirical} depicts a comparison of the RMS Doppler spread obtained by \eqref{eq:analyitcalDopplerSpread} and the RMS Doppler spread obtained from simulations using the channel model described in Sec.~\ref{subsec:TDLModel} The delay parameter $\tau_0$ is varied from $34\,\text{ns}$ to $150\,\text{ns}$ and $f_{\text{LOS}} \in\{ 0, 250\}\,\text{Hz}$. We can see in Figure~\ref{fig:DopplerSpreadAnalyticalvsEmpirical} that these two RMS Doppler spread values are very close, concluding that \eqref{eq:analyitcalDopplerSpread} gives an enough accurate representation of the RMS Doppler spread. From \eqref{eq:analyitcalDopplerSpread} it is evident that the Doppler shift of the LOS component, the $K$-factor and the power of the delay taps influence the RMS Doppler spread. We further see that for $K>0$ (first tap Rician distributed) the relation between $f_\text{Dmax}$ and $\sigma_\nu$ is more complex, with respect to the other taps which exhibit Rayleigh fading only. 

Furthermore, we can see that the power of the delay taps $\alpha[i]$ also influences the RMS Doppler spread. The power of the delay taps clearly depends on the delay parameter, which can be seen in \eqref{pdpdiscrete}, and thus on the RMS delay spread. Hence, the RMS delay spread has an impact on the obtained RMS Doppler spread. In Figure~\ref{fig:DopplerSpreadAnalyticalvsEmpirical} we see that for an RMS delay spread of approx. $25\,\text{ns}$ and a $K$-factor of $20\,\text{dB}$, we obtain an RMS Doppler spread of approximately $100\,\text{Hz}$ (whereas $f_\text{Dmax}=500\,\text{Hz}$). Increasing the RMS delay spread results in an increase of the RMS Doppler spread. This increase is non-linear and its steepness depends on the Doppler bandwidth and $K$-factor.

\begin{figure}[t!]    
	\centering
	\includegraphics[width=0.95\columnwidth]{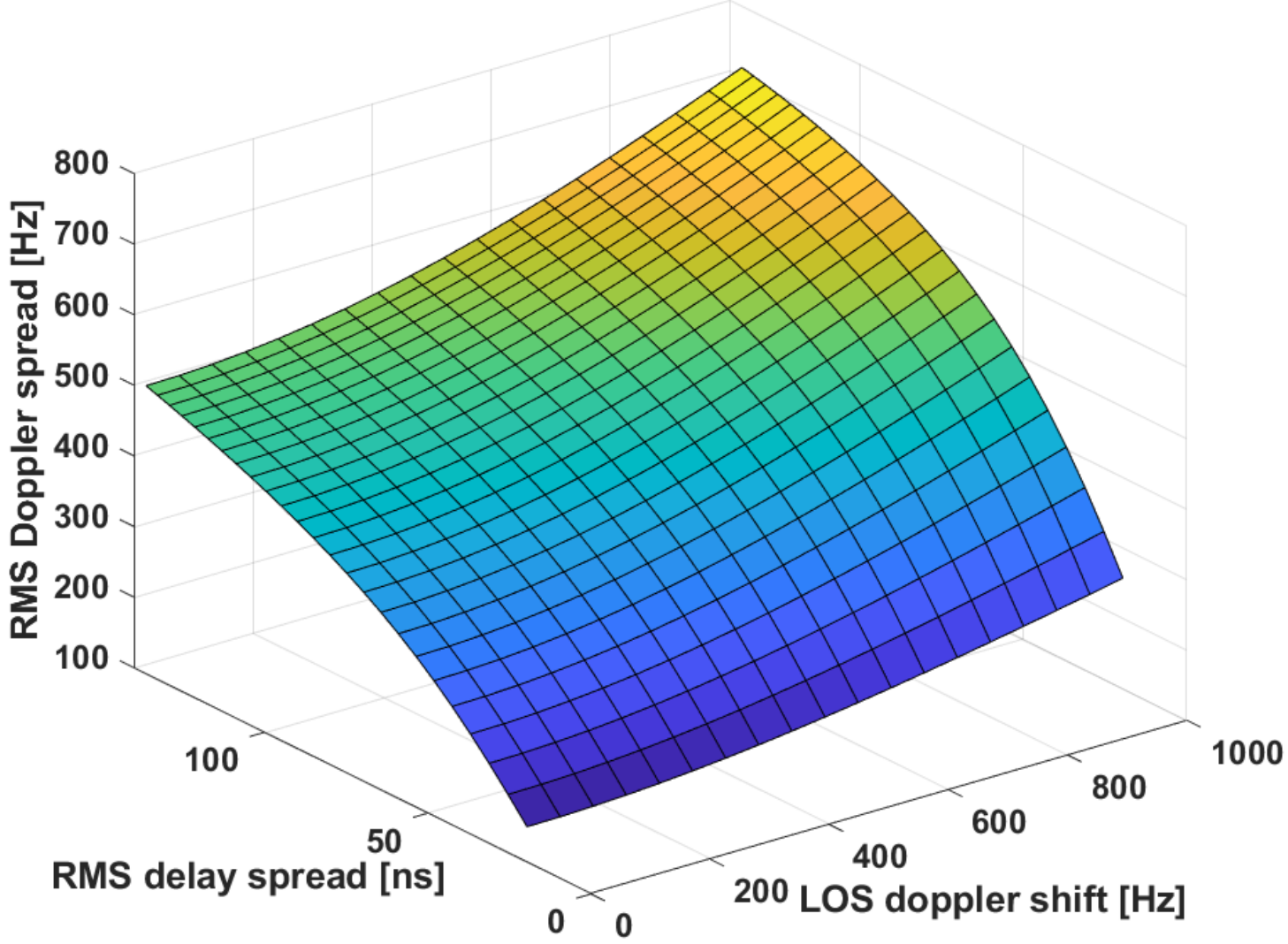}
	\caption{Analytical Doppler spread for a fixed $K$-factor of $20\,\text{dB}$ and $f_\text{Dmax} = 1\,\text{kHz}$, for different values of RMS delay spread and Doppler shift of the LOS component}
	\label{fig:K20FDmax2kHz}
\end{figure}

Using \eqref{eq:analyitcalDopplerSpread} we are now able to investigate how the RMS Doppler spread changes if a strong LOS component is present with respect to a given RMS delay spread and LOS Doppler shift. First, we see already in Figure~~\ref{fig:DopplerSpreadAnalyticalvsEmpirical} that for the fixed RMS delay spread and $K$-factor the RMS Doppler spread increases with increasing the LOS Doppler shift, by showing it for $f_{\text{LOS}} = 0\,\text{Hz}$ and $f_{\text{LOS}} = 250\,\text{Hz}$. In Figure~ \ref{fig:K20FDmax2kHz} we investigate this influence further and show the results. Figure~ \ref{fig:K20FDmax2kHz} depicts this relation for $K=20\,\text{dB}$ and $f_\text{Dmax}=1\,\text{kHz}$. The influence of the LOS Doppler shift on the RMS Doppler spread clearly depends on the RMS delay spread. We further see an increase of the RMS Doppler spread by approximately $200\,\text{Hz}$ when the LOS Doppler shift changes from $0$ to $1\,\text{kHz}$. 

\subsection{Measurement Setup}
\label{subsec:meassetup}

\begin{figure}[t!]    
	\centering
	\includegraphics[width=0.95\columnwidth]{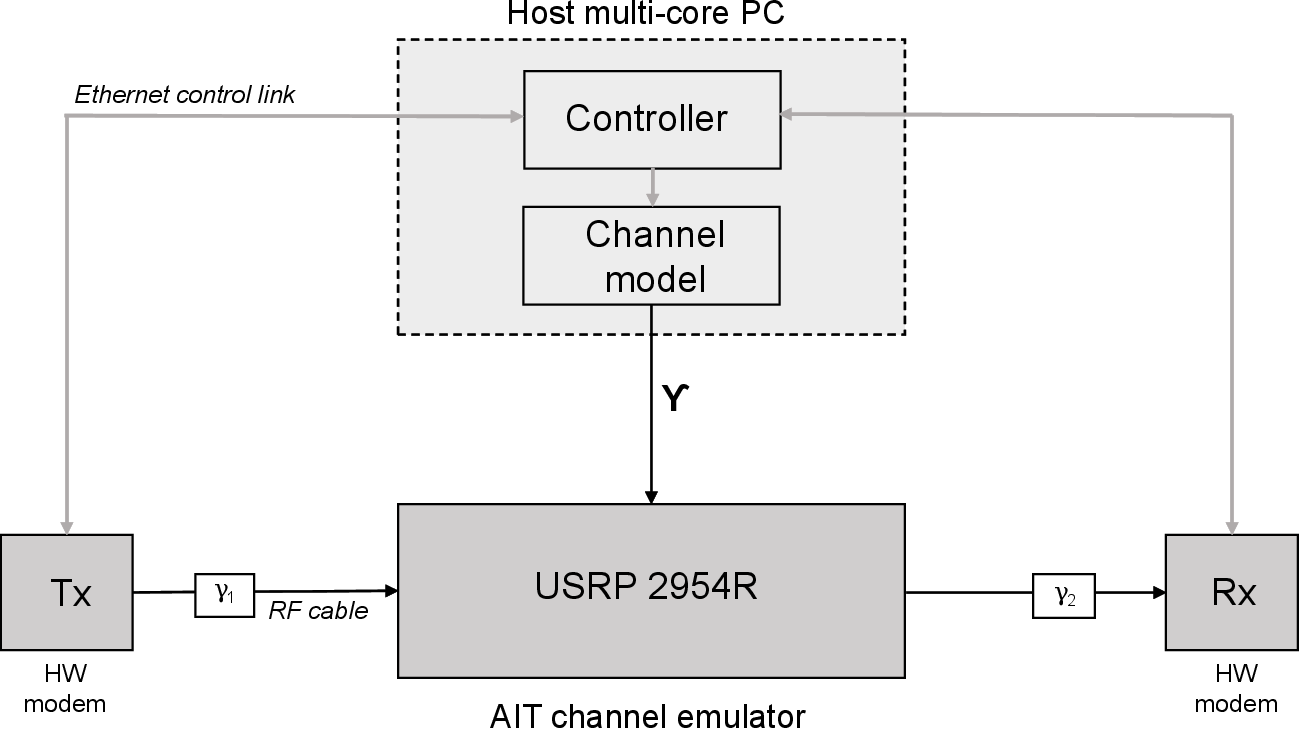}
	\caption{Setup of the FER measurement. $\Upsilon$ presents the compressed form of the channel impulse response.}
	\label{fig:systemmodel}
\end{figure}

\begin{figure}[t!]    
	\centering
	\includegraphics[width=0.99\columnwidth]{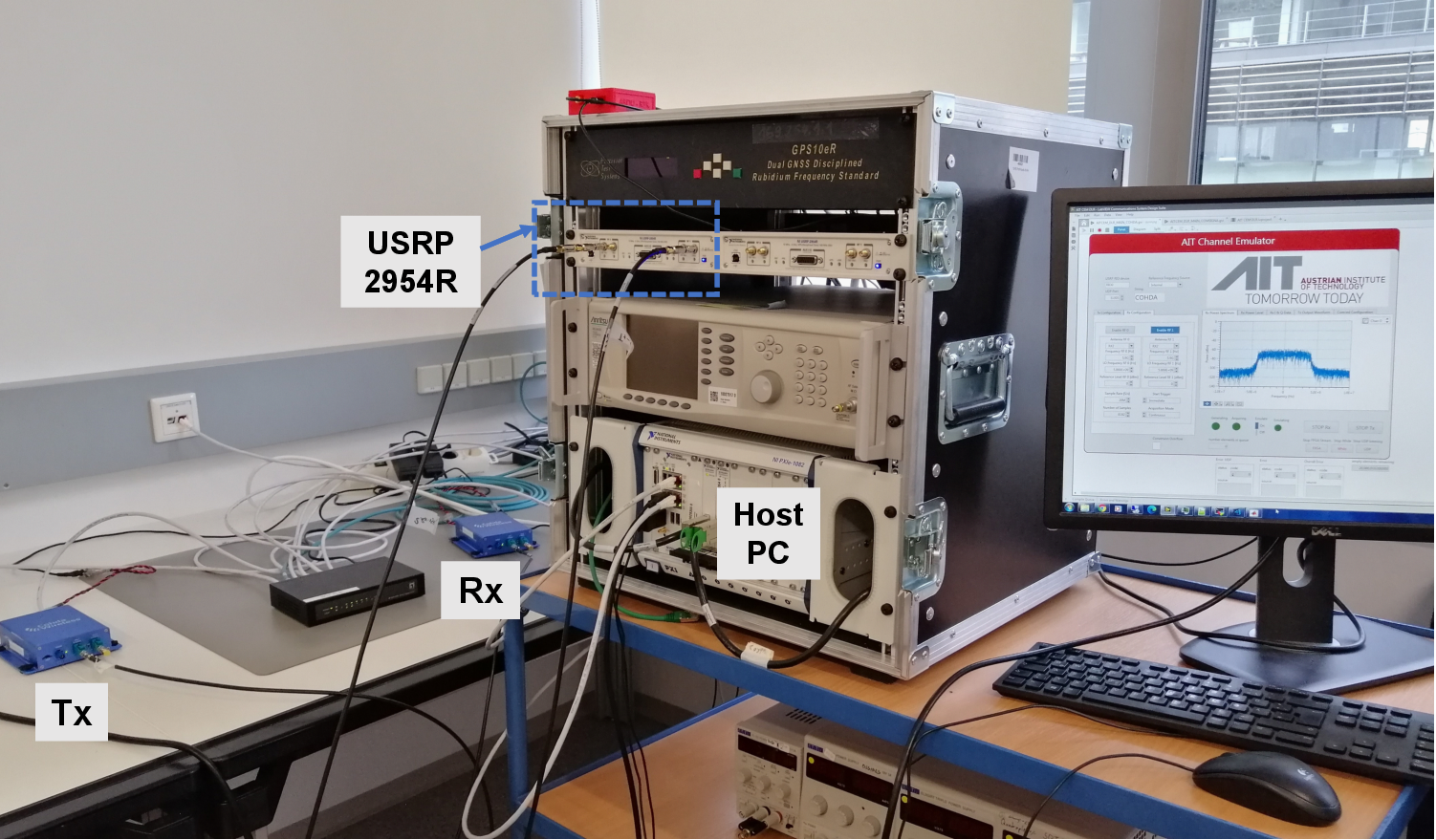}
	\caption{Measurement setup in the laboratory environment.}
	\label{fig:lab}
\end{figure}

The measurement setup to obtain the FER lookup table entries is presented in  Fig.~\ref{fig:systemmodel}. 
We measure the FER for the condensed radio channel parameter vector $\Vec{\Psi}$. Our methodology is generally applicable to any communication standard (4G, 5G, IEEE 802.11x) or modem type. As an example in this paper, we use Cohda Wireless MK5 on-board unit modems \cite{Cohda} implementing the IEEE 802.11p PHY. The Tx modem is connected via a radio frequency (RF) cable to the input of the AIT real-time channel emulator \cite{Hofer19}, attenuated by $\gamma_1 = 26\,\text{dB}$. The output of the emulator is connected to the Rx modem, attenuated by $\gamma_2 = 56\,\text{dB}$. The AIT real-time channel emulator consists of a software defined radio (SDR), in our case the National Instrument USRP 2954R, and a multi-core personal computer (PC) that implements the channel model. On the SDR a field programmable gate array (FPGA) convolves the input signal from the Tx modem with the doubly selective channel impulse response to obtain the output signal that is fed to the Rx modem. The channel impulse response is obtained in compressed form ($\boldsymbol{\Upsilon}$ in Fig.~\ref{fig:systemmodel}) from the channel model on the host PC (for a detailed explanation see \cite{Hofer19}) and is transferred to the FPGA where the channel is emulated. From the multi-core PC we control the system using a Python script, which is presented as controller in Fig.~\ref{fig:systemmodel}. From the controller, using a secure shell (SSH) connection, we access the Tx and Rx modems, configure them and run transmission (Tx modem) and reception (Rx modem) commands. The measurement setup used in the laboratory is depicted in Fig.~\ref{fig:lab}.

\subsection{Measurement Process}
\label{subsec:measproc}

The measurement process is started from the controller. The FER is obtained by repeating the frame transmission $F$ times for all condensed channel parameter sets $\Vec{\Psi}\in \mathcal{I}$. The channel model parameter set $\mathcal{I}$ is shown in Table~\ref{table2}. To obtain different received power values we change the channel attenuation, while the transmit power $P_\text{Tx}$ and the attenuation between Tx and Rx modem, which are caused by the cables, the physical attenuators and the USRP, stay fixed. Hence, we calculate the power on the receiver side using \eqref{eq:Rxpower}, where the total path loss $\rho$ is variable. In each measurement iteration, we firstly establish an SSH connection with the Tx and Rx modem and configure them. Secondly, we run the receiving command on the Rx modem and execute the channel model with a certain path loss, RMS delay spread, Doppler bandwidth, $K$-factor and LOS Doppler shift combination $\Vec{\Psi}$. Thirdly, when the channel emulator has started, we run the transmit command on the Tx modem with the parameters shown in Table~\ref{table}.
The number of successfully received frames after each repetition is saved on the Rx modem. The FER is calculated as a ratio of the number of frames received with errors over the total number of transmitted frames. We transmit $F=5\cdot 10^6$ frames for each parameter combination.

\begin{table}
	\begin{center}
	\caption{Transmission and communication parameters.}
	\label{table}
	\setlength{\tabcolsep}{2pt}
	\begin{tabular}{|p{150pt}|p{75pt}|}
		\hline 
		{\small Parameter}& 
	{\small Value}\\
		\hline
		{\small Number of transmitted frames, $F$}& 
		{\small$5 \cdot 10^6$}\\
		{\small  Frame size}& 
		{\small $100$ bytes}\\
		{\small Frame rate}& 
		{\small $2200\,\text{packets/s}$} \\
		{\small Transmit power, $P_\text{Tx}$}& 
		 {\small $-5\,\text{dBm}$}  \\
		{\small Modulation and coding rate}& 
		{\small QPSK; $C=1/2$}\\
		{\small Number of delay taps, $N_ \text{taps}$}& 
		{\small $8$}\\
		{\small Resolution of taps, $\Delta t$}& 
		{\small $100\,\text{ns}$} \\
		{\small Time delay paths in every tap, $L'$}& 
		{\small $40$} \\
		{\small Carrier frequency, $f_\text{C}$}& 
		{\small $5.9\,\text{GHz}$} \\
		{\small Communication bandwidth, $B$}& 
		{\small $10\,\text{MHz}$} \\
		{\small Sampling delay time, $T_\text{c}$}& 
		{\small $10\,\text{$\mu$s}$} \\
		\hline
	\end{tabular}
    \end{center}
\end{table}

\begin{table}
	\begin{center}
	\caption{Channel model parameters discretization used for the FER lookup table measurements.}
	\label{table2}
	\setlength{\tabcolsep}{1pt}
	\begin{tabular}{|p{75pt}|p{150pt}|}
		\hline 
		{\small Parameter}& 
	{\small Discrete value set}\\
		\hline
		{\small RMS delay spread }& 
		{\small $I_{\sigma_\tau}=\{25, 50, 82 \}\,\text{ns}$}\\
		{\small Doppler bandwidth}& 
		{\small $I_{f_\text{Dmax}}=\{10, 50, 100, 500, 1000\} \,\text{Hz} $}\\
		{\small $K$-factor}& 
		{\small $I_K=\{-\infty, 10, 15, 20\}\, \text{dB} \,$ }\\
		{\small LOS Doppler Shift}&
		{\small $I_{f_\text{LOS}}=\{0, f_\text{Dmax} / 2, f_\text{Dmax}\}$ \, }\\
		{\small Received power} & 
		 {\small $I_P=\{-94.9,-92.9 . . .-78.9\}\,\text{dBm}$}  \\
		\hline
	\end{tabular}
    \end{center}
\end{table}

\subsection{Resolution Analysis}

\begin{figure}[t!]    
	\centering
	\includegraphics[width=0.95\columnwidth]{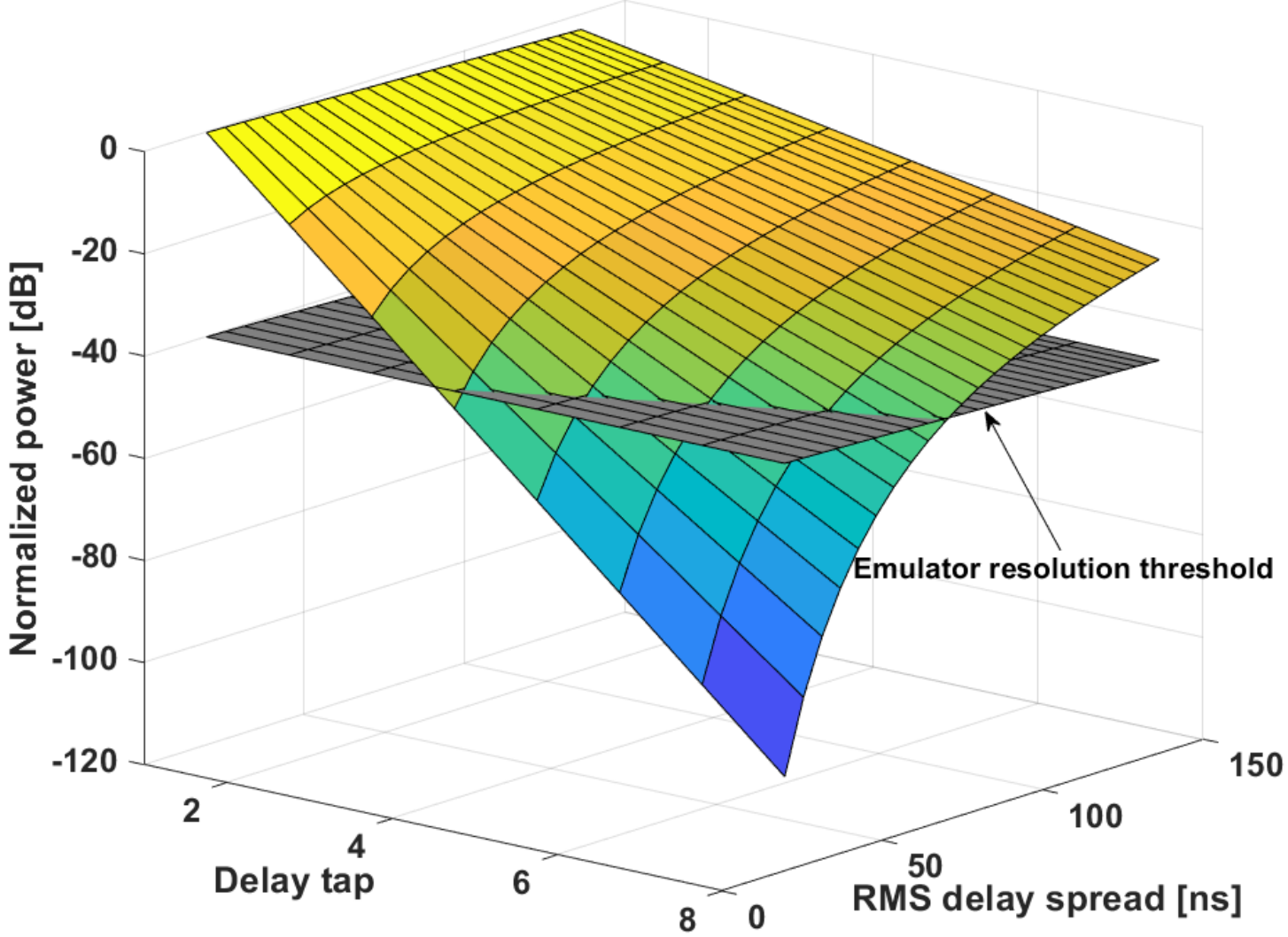}
	\caption{Normalized power for each delay tap and a given RMS delay spread according to \eqref{eq:normalization2}. The grey plane indicates the emulator resolution threshold/dynamic range.}
	\label{fig:normalizedPowerPerTabVsDelaySpread}
\end{figure}

\begin{figure}[t!]    
	\centering
	\includegraphics[width=0.95\columnwidth]{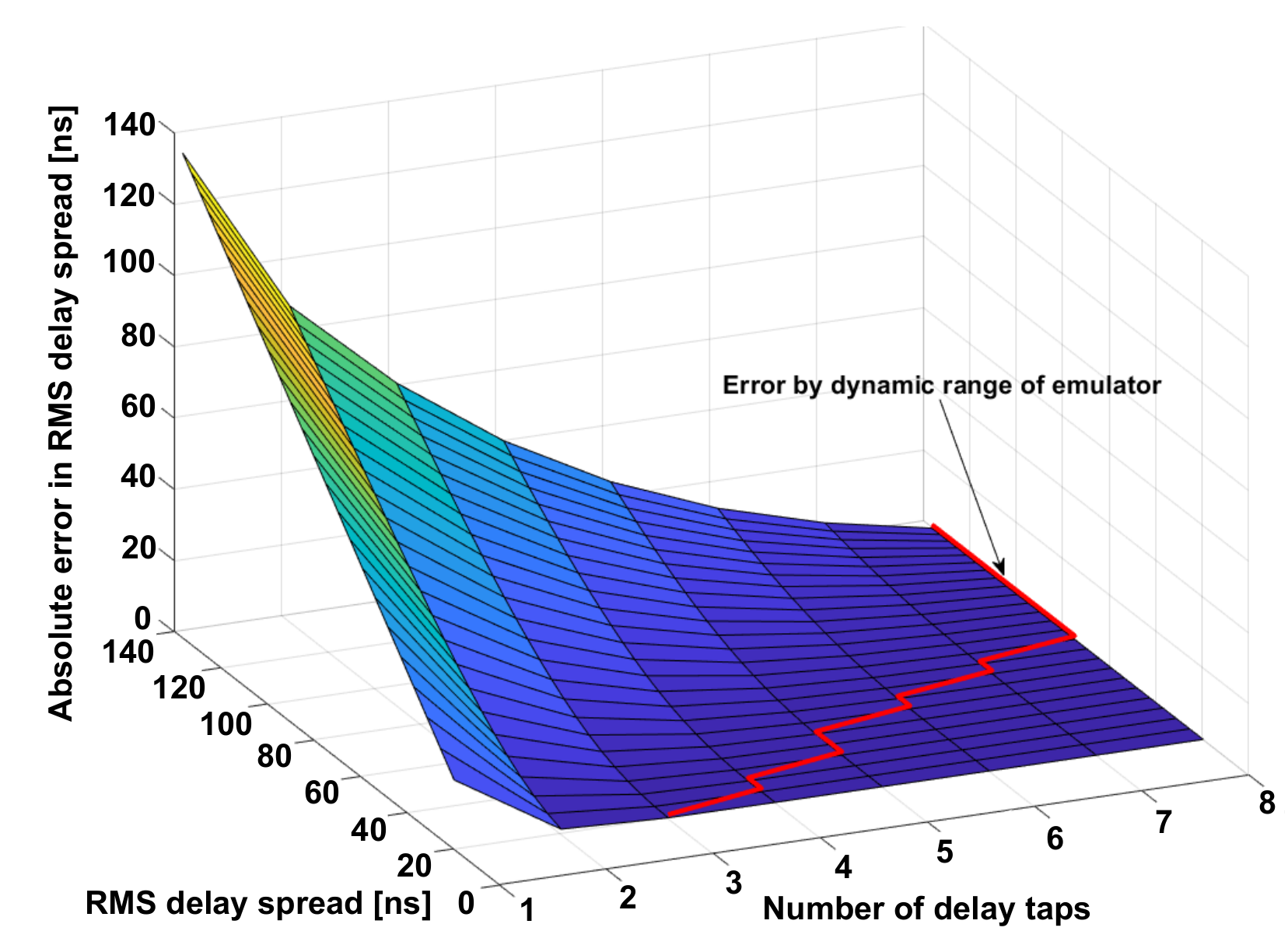}
		\caption{Absolute error of resulting RMS delay spread given a targeted RMS delay spread and a certain number of considered taps. The red line denotes the error introduced by the resolution threshold/dynamic range of the emulator.}
		\label{fig:delaySpreadErrorByDynamicRangeofEmu}
	\end{figure}

In order to show the validity and applicability of the chosen stochastic channel model and the selection of its parameters, we investigate the impact of the dynamic range of the emulator. Figure~\ref{fig:normalizedPowerPerTabVsDelaySpread} depicts the normalized average power for each considered delay tap and RMS delay spreads starting at $20\,\text{ns}$. The grey plane indicates the resolution threshold of the AIT channel emulator. This indicates that for lower RMS delay spreads the emulator will most likely not be able to resolve the channel for higher delay taps and, thus, may reduce the actual RMS delay spread the receiver will be able to witness. In Figure~\ref{fig:delaySpreadErrorByDynamicRangeofEmu} we show the absolute error in RMS delay spread obtained when the emulator uses a different number of taps (x-axis) to obtain a target RMS Delay spread (y-axis) i.e., either one, two, or more taps are used to obtain the targeted delay spread. 
The red line denotes the error introduced by the resolution threshold, corresponding to the intersecting planes in Figure~\ref{fig:normalizedPowerPerTabVsDelaySpread}. The RMS delay spread and number of taps combinations that are on the right hand side of this line, are not possible to be emulated due to the dynamic range limitation. This limitation has an negligible impact on the resulting RMS delay spread of the emulated channel. The highest error caused by the limitations of the emulator's dynamic range is approximately $0.1\,\text{ns}$ and it is attained for the lowest RMS delay spread of $20\,\text{ns}$. Thus, we conclude that the emulator is able to accurately represent the generated channels with the parameters given in Table~\ref{table2}.

\section{System-Level Simulation}
\label{sec:SLS}
In Figure~\ref{fig:process} (b) we show the structure of our system-level simulation. The system-level simulation is based on a GSCM \cite{Karedal09} and a FER lookup table, whose entries are obtained as described in Sec.~\ref{sec:LUTgenerator}. We use a small number of condensed radio channel parameters to index the FER lookup table. These condensed radio channel parameters are directly estimated in real-time from the time-variant propagation path parameters of the GSCM, establishing a link between the GSCM and the FER lookup table. 

\subsection{Geometry-based Stochastic Channel Model}
The channel model differentiates between the following contributions: LOS between Tx and Rx, reflections from static discrete scatterers (SD), reflections from moving discrete scatterers (MD) and diffuse scattering (D). The geometry for the channel model is imported from OpenStreetMap following the approach outlined in \cite{Rainer20}. The discrete time-variant frequency response is calculated by
\begin{multline}
 H[m', k] = \eta^{\text{LOS}}[m'] \me^{-\mj 2 \pi k \Delta f \tau^{\text{LOS}}[m']} + \\
+ \sum_{i=1}^{N_{\text{SD}}} \eta_{i}^{\text{SD}}[m'] \me^{-\mj 2 \pi (f_\text{C} + k \Delta f) \tau^{\text{SD}}_{i}[m']} +  \\
+ \sum_{i=1}^{N_{\text{MD}}} \eta_{i}^{\text{MD}}[m'] \me^{-\mj 2 \pi (f_\text{C} + k \Delta f) \tau^{\text{MD}}_{i}[m']} + \\
+ \sum_{i=1}^{N_{\text{D}}} \eta_{i}^{\text{D}}[m'] \me^{-\mj 2 \pi (f_\text{C} + k \Delta f) \tau^{\text{D}}_{i}[m']}, 
\end{multline}
where discrete frequency $k \in \{ -\lfloor \frac{B}{2\Delta f} \rfloor, \ldots, \lfloor \frac{B}{2\Delta f} \rfloor - 1 \}$, $f_\text{C}$ denotes the carrier frequency and $ \Delta f $ denotes the subcarrier spacing. 

Diffuse scatterers are automatically distributed along the buildings, according to \cite{Rainer20}. Tx, Rx and mobile discrete scaterrer trajectories are represented using GPS coordinates. The trajectories are then interpolated in time using the modified Akima interpolation \cite{Akima70} with the stationarity region duration as supporting point distance. 

\subsection{Low-complexity estimation of condensed radio channel parameters}
\label{subsec:Computationparameters}
We are interested in efficiently estimating five condensed radio channel parameters within each stationarity region $r$ directly from the propagation paths obtained by a GSCM. These parameters are: the received power $P_r$, the RMS delay spread $\sigma_{\tau,r}$, the Doppler bandwidth $f_{\text{Dmax},r}$, the $K$-factor of the LOS delay tap $K_r$, and the Doppler shift of the LOS path $f_{\text{LOS},r}$. These five parameters are the indices of our FER lookup table. We also show the RMS Doppler spread $\sigma_{\nu,r}$ estimation, since it is used for analyzing the FER results.

\subsubsection{RMS Delay Spread}
\label{subsubsec:RMSdelayspread}
We can obtain the short-term PDP by averaging the channel impulse response from \eqref{discereteCIR} over $M$ time samples for each stationarity region $r\in \left\lbrace 0, \ldots, R-1\right\rbrace$ as
\begin{equation}
\label{eq:PDP1}
P_{\tau,r}[n] =\frac{1}{M}\sum_{m=0}^{M-1} \left|h[rM+m,n])\right|^2\,.
\end{equation}

However, to reduce complexity we want to compute the PDP directly from the propagation path parameters avoiding the explicit computation of the time-variant channel impulse response. Starting from \eqref{discereteCIR2} we obtain
\begin{multline}
    \label{eq:PDP3_infM}
	P_{\tau,r} [n] \approx  \\ \approx\frac{1}{M}\sum_{m=0}^{M-1}\left\lvert \sum_{l=1}^L \left\lvert\eta_{l,r}\right\rvert\me^{\mj2\pi(\phi_{l,r}-\nu_{l,r}m))}h_\text{RC}(nT_\text{c}-\tau_{l,r})\right\rvert^2 =\\
	=\frac{1}{M}\sum_{m=0}^{M-1}\left\lvert\sum_{l=1}^L a_{l,r}[n]\me^{-\mj2\pi\nu_{l,r}m)}\right\rvert^2 = \\
	=\sum_{l=1}^L\lvert a_{l,r}[n]\rvert^2 + \frac{1}{M}\sum_{m=0}^{M-1}\sum_{l=1}^L \sum_{k \neq l}^{L}a_{l,r}[n]a_{k,r}[n]^* \\
	\cdot \me^{-\mj2\pi\nu_{l,r}m)}\me^{\mj2\pi\nu_{k,r}m)},
\end{multline}
where $a_{l,r}[n] := |\eta_{l,r}|e^{j2\pi \phi_{l,r}}h_\text{RC}(nT_\text{c}-\tau_{l,r})$ and $a_{k,r}[n] := |\eta_{k,r}|e^{j2\pi \phi_{k,r}}h_\text{RC}(nT_\text{c}-\tau_{k,r})$. We proceed further with the computation of the mixed terms from \eqref{eq:PDP3_infM}, which is shown in detail in Appendix \ref{app:PDPapprox}, and obtain 
\begin{multline}
\label{eq:PDPSS}
P_{\tau,r}[n]=\sum_{l=1}^{L}\left\lvert\eta_{l,r}\right\rvert^2\left\lvert h_\text{RC}(nT_\text{c}-\tau_{l,r})\right\rvert^2 \\ 
+\frac{1}{M} \sum_{l=1}^L\sum_{k>l}^L \left\lvert\eta_{l,r}\right\rvert \left\lvert\eta_{k,r}\right\rvert h_\text{RC}[n -\tau_{l,r}]h_\text{RC}[n -\tau_{k,r}] \cdot \\
\cdot 2
\begin{cases}
\sin\left(\frac{\theta_{l,k,r}}{2}\right)^{-1}\sin\left(\frac{\theta_{l,k,r}}{2}M\right) \cos(\Psi_r) \,\, \text{for}\,\,\, \theta_{l,k,r}\neq 0;\\
M \cos\left(2\pi(\phi_{l,r} - \phi_{k,r})\right) \quad \text{otherwise}\,;
\end{cases}
\end{multline} 
with $\Psi_r =  \frac{\theta_{l,k,r}}{2}(1-M) - 2\pi(\phi_{l,r} - \phi_{k,r})$ and $\theta_{l,k,r} = 2\pi(\nu_{k,r} - \nu_{l,r})$.

The computational complexity of \eqref{eq:PDPSS} does no longer depend on the number of time samples $M$ but only on the propagation path parameters. In comparison to \eqref{eq:PDP1} which has complexity $O(M(L + L^2))$ we have now a complexity of $O(L+L^2/2)$. The trigonometric functions can also be computed in a fast manner using fixed point lookup tables.

The RMS delay spread is the square root of the second order central moment $\mathbb{E}[(X-\mathbb{E}[X])^2]$ of the PDP \cite{Hlawatsch11}
\begin{equation}
\sigma_{\tau,r} = \sqrt{\frac{\sum_{n=0}^{N-1}(nT_\text{c})^2P_{\tau,r}[n]}{\sum_{n=0}^{N-1}P_{\tau,r}[n]}-(\overset{-}{\tau_r})^2}
\label{delayspread}
\end{equation}
where $\overset{-}{\tau_r}$ is the mean delay calculated as the normalized first order moment
\begin{equation}
\overset{-}{\tau_r} = \frac{\sum_{n=0}^{N-1}(nT_\text{c})P_{\tau,r}[n]}{\sum_{n=0}^{N-1}P_{\tau,r}[n]} .
\label{meandelay}
\end{equation}
\subsubsection{RMS Doppler Spread}
\label{DopplerShiftFreq}
Computing the discrete Fourier transform \cite{Rao00} of the time-variant channel impulse response  \eqref{discereteCIR}, taking \eqref{eq:ConstVel} and \eqref{eq:ConstWeight} into account, we obtain the Doppler-variant impulse response for stationarity region $r$,
\begin{multline}
	s_r[p,n] \approx \sum_{l=1}^{L} |\eta_{l,r}| \me^{\mj 2 \pi \phi_{l,r}} h_\text{RC}(nT_\text{c}-\tau_{l,r})  \cdot \\
	\cdot \text{sinc}\left( \left(f_{l,r} -\frac{p}{T_{\text{stat}}}\right)T_{\text{stat}}\right)
	\label{discereteCTF}.
\end{multline}
where $p \in \left\lbrace -\frac{M}{2},...,\frac{M}{2}-1\right\rbrace $ represents the normalized discrete Doppler index. A detailed derivation is shown in Appendix \ref{app:DSDapprox}. 

In order to obtain the short-term DSD,  $s[p,n]$ is averaged over the delay
\begin{multline}
    P_{\nu,r}[p]= \frac{1}{N}\sum_{n=0}^{N-1}\left\lvert \sum_{l=1}^{L} |\eta_{l,r}| \cdot\right. \\
\left. \cdot \me^{\mj 2 \pi \phi_{l,r}} h_\text{RC}(nT_\text{c}-\tau_{l,r})\text{sinc}( u_{l,r}[p] )\vphantom{\sum_{l=1}^{L}} \right\rvert ^2
\label{averageCTF}
\end{multline}
with $u_{l,r}[p]:=\left(f_{l,r} -\frac{p}{T_{\text{stat}}}\right)T_{\text{stat}}$. Similar to the RMS delay spread, we obtain the RMS Doppler spread \cite{Hlawatsch11}
\begin{equation}
\sigma_{\nu,r} = \sqrt{\frac{\sum_{p=0}^{M-1}(pT_s)^2P_{\nu,r}[p]}{\sum_{p=0}^{M-1}P_{\nu,r}[p]}-(\overset{-}{\nu_r})^2},
\label{Dopplerspread}
\end{equation}
where $\overset{-}{\nu_r}$ is the mean Doppler shift calculated as the normalized first order moment
\begin{equation}
\overset{-}{\nu_r} = \frac{\sum_{p=0}^{M-1}(pT_s)P_{\nu,r}[p]}{\sum_{p=0}^{M-1}P_{\nu,r}[p]} .
\label{meanDoppler}
\end{equation}
\subsubsection{Doppler Bandwidth}
\label{DopplerBW}
We use the approximated DSD in \eqref{averageCTF} for determining the approximate Doppler bandwidth. The Doppler bandwidth $f_{\text{Dmax},r}$ is then approximated by thresholding \eqref{averageCTF} using a specific dynamic range $\varepsilon$ (i.e., $\varepsilon=40$ dB). The frequencies for which $P_{\nu,r}[p]$ is greater than $\max{P_{\nu,r}[p]} / 10^{\varepsilon/10}$ are frequencies which remain. Finally, $f_{\text{Dmax},r}$ is calculated as a difference between maximum and minimum value of all remaining frequencies.

\subsubsection{Rician $K$-Factor}
\label{$K$-factor}
In the case of non-obstructed LOS, we estimate the $K$-factor by the ratio of the power of the LOS component and the sum of the power of scattered components which fall into the LOS delay bin outlined in \eqref{eq:KFactorPaths} for the $m$-th stationarity region sample in our GSCM.
\begin{equation}
\label{eq:KFactorPaths}
K_r = \frac{|\eta_{\text{LOS},r}|^2}{|\sum_{l \in L_r\setminus \{\text{LOS}\}} \eta_{l,r} \delta_\text{K}(\lfloor\tau_{l,r} B\rfloor - \lfloor\tau_{\text{LOS},r} B \rfloor )|^2}
\end{equation}
where $L_r$ denotes the set of all paths impinging at the receiver in different delay bins in stationarity region $r$. In case that there is only LOS present in its delay bin we set the $K$-factor to $500\,\text{dB}$. In NLOS situations we set $K_r = 0$.

\subsubsection{Received Power}
\label{ReceivedPower}
To obtain the received power $P_r$ we apply a threshold of $40~\text{dB}$ and calculate the path loss as  the integrated power of our estimated instantaneous PDP in (\ref{eq:PDPSS}).

\subsubsection{LOS Doppler Shift}
\label{LOSDopplerShift}
The LOS Doppler shift $f_{\text{LOS},r}$ for stationarity region $r$ is directly available from the GSCM LOS propagation path.

\section{Results and discussion}
\label{sec:resanddiscusion}
The FER lookup table provides the system-level simulator with the FER for each stationarity region during the run-time. It has to be pre-measured, which requires a certain amount of time. The used hardware modem, frame sizes, modulation and coding rate, determines the data throughput. To obtain reliable FER measurements we need to define a minimum FER $\kappa= 2\cdot 10^{-5}$ and variance of the FER estimates $\iota=0.01$. With these parameters we can define the amount of frames per data point as $F=1/(\kappa \iota)= 5\cdot 10^{6}$ frames.


\subsection{FER Results of V2X Modems}
\label{sec:results}
We measured the FER for $1800$ parameter combinations of the condensed channel parameter vector $\Vec{\Psi}$, shown in Table~\ref{table2}, to fill the FER lookup table using the automated procedure described in Sec. \ref{sec:LUTgenerator}. Here a selected subset of these measurement results is presented. 

A quadrature phase shift keying (QPSK) modulation and coding rate of $C=1/2$ are used, respectively. The Rx modem report the noise power $P_N$ and the signal power $P$ for each received data frame. We show FER results with respect to the SNR $\zeta$ calculated as
\begin{equation}
\zeta =P-P_N,
\end{equation}
averaging over all $F$ transmitted frames.

\subsubsection{$K$-Factor Variation}
\begin{figure}[!h]    
	\centering
	\includegraphics[width=0.95\columnwidth]{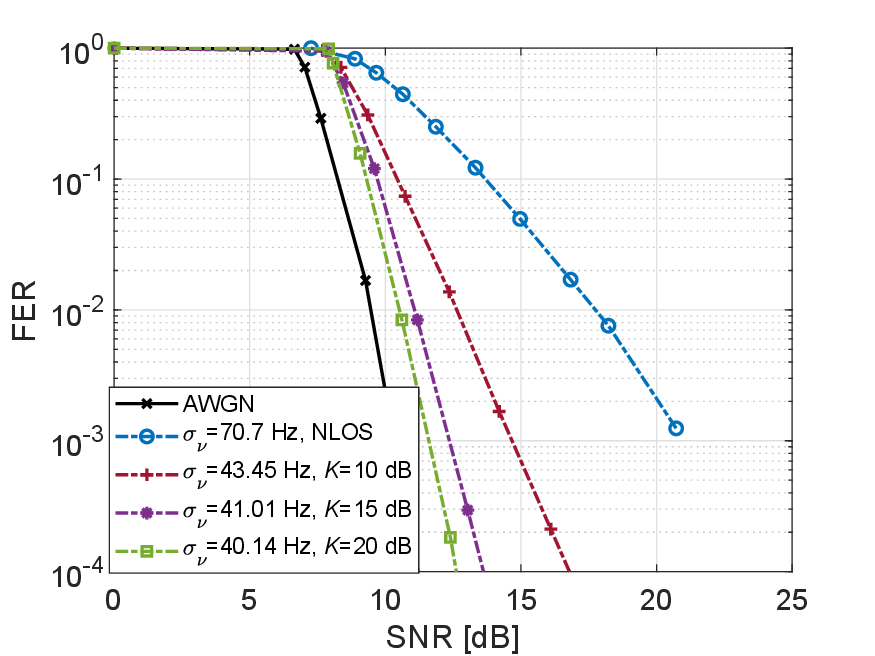}
	\caption{Measurement results of FER vs. SNR using an exponential PDP with $K$-factors $K\in\{-\infty,10,15,20\}\,\text{dB}$, RMS delay spread of $82\,\text{ns}$ and Doppler bandwidth $f_\text{Dmax}= 100\,\text{Hz}$, having $f_{\text{LOS}} = 0\,\text{Hz}$.}
	\label{fig:kfactors}
\end{figure}

In Figure~\ref{fig:kfactors} we show the FER measurement results for different $K$-factors with an RMS delay spread of $82\,\text{ns}$ and a Doppler bandwidth of $500\,\text{Hz}$. The highest FER is obtained for the NLOS condition \cite{Bernado15}. Increasing the $K$-factor, the LOS component becomes more dominant. This results in lower probability of deep fades. 
Therefore, the FER decreases with higher $K$-factor, as we see in Figure~\ref{fig:kfactors}.

\subsubsection{RMS Delay Spread Variation}
\begin{figure}[!t]    
	\centering
	\includegraphics[width=0.95\columnwidth]{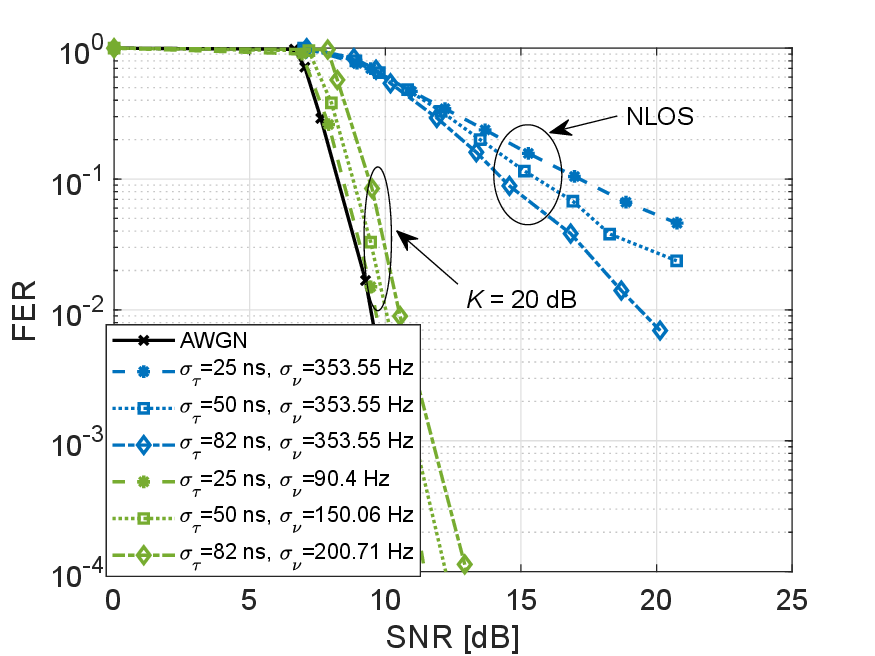}
	\caption{Measurement results of FER vs. SNR using an exponential PDP with  RMS delay spread $\sigma_\tau\in\{25,50,82\}\,\text{ns}$, $K$-factors $K\in\{-\infty,20\}\,\text{dB}$ and Doppler bandwidth $f_\text{Dmax} = 500\,\text{Hz}$, having $f_{\text{LOS}} = 0\,\text{Hz}$.}
	\label{fig:delayspreads}
\end{figure}

Furthermore, we investigate how the FER changes for different RMS delay spread values in the case of LOS and NLOS. In Figure~\ref{fig:delayspreads} we show the measurement results for the RMS delay spreads of $25,\, 50\, \text{and}\, 82\,\text{ns}$ for NLOS and $K=20\,\text{dB}$ in LOS. The Doppler bandwidth used for these measurements is $500\,\text{Hz}$. 

First of all, we notice a better Rx performance in terms of FER when we consider a LOS component.
Then we observe that if the RMS Doppler spread is constant (as in the NLOS case in Figure~\ref{fig:delayspreads}), the FER decreases with increasing RMS delay spread. However, we observe the opposite effect when we introduce a LOS component with a given $K$-factor. Then the FER increases with increasing RMS delay spreads. In this case, the Doppler spread will be influenced by the RMS delay spread, as shown in Figure~\ref{fig:DopplerSpreadAnalyticalvsEmpirical}, in a way that if RMS delay spread increases, the RMS Doppler spread also increases. 
 
In the NLOS case, the receiver will exploit the channel diversity in the delay domain and deliver better FER results when the RMS delay spread increases.
On the other hand, in the LOS case, with $K=20\,\text{dB}$ the measurements results show that the Rx cannot cope with the spread in the Doppler domain. When the RMS delay spread gets larger, consequently also the RMS Doppler spread (Figure~\ref{fig:DopplerSpreadAnalyticalvsEmpirical}) increases, and the FER results get worse. Based on that, we can conclude that the RMS Doppler spread has a stronger impact on the FER than the RMS delay spread.

\subsubsection{Doppler Bandwidth Variation}
\begin{figure}[!t]    
	\centering
	\includegraphics[width=0.95\columnwidth]{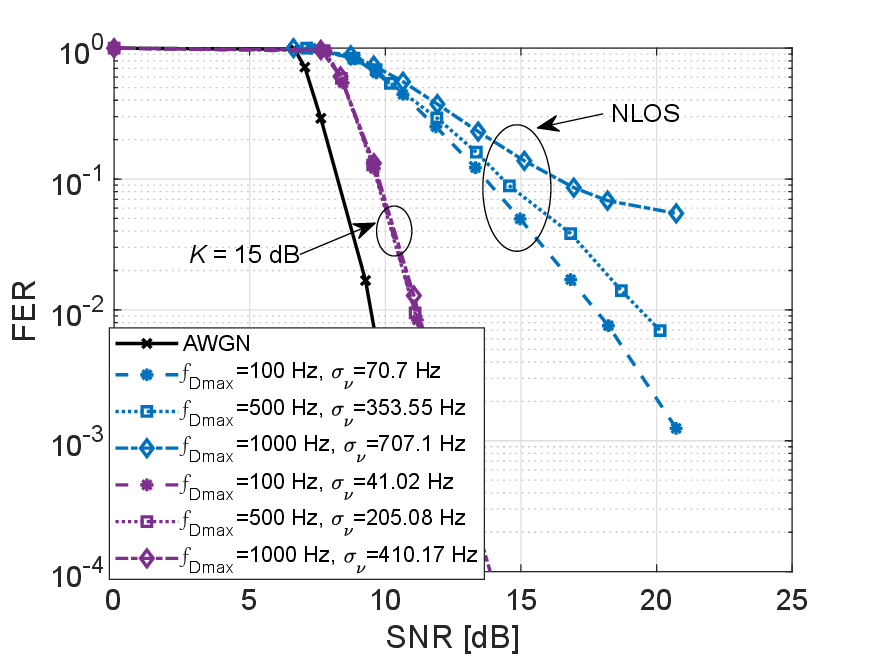}
	\caption{Measurement results of FER vs. SNR using an exponential PDP with  Doppler bandwidth $f_\text{Dmax}\in\{100,500,1000\}\,\text{Hz}$, $K$-factors $K\in\{-\infty,15\}\,\text{dB}$ and RMS delay spread $\sigma_ \tau = 82\,\text{ns}$, having $f_{\text{LOS}} = 0\,\text{Hz}$.}
	\label{fig:dopplerspreads}
\end{figure}

In Fig.~\ref{fig:dopplerspreads} we show the FER performance for the different Doppler bandwidth values. In the NLOS case the best FER performance is obtained for the Doppler bandwidth of $100\,\text{Hz}$, which is the lowest one. Increasing the Doppler bandwidth leads to higher RMS Doppler spreads and higher FERs. The reason is, that a higher Doppler bandwidth results in a faster change of the channel, which in turn, leads to increased channel estimation errors \cite{Molisch10}. This is a well known limitation of the IEEE 802.11p pilot pattern as discussed in detail in \cite{Zemen12, Zemen12a, Mecklenbrauker11}. In the LOS case, due to the strong LOS component, we see no impact of the different Doppler bandwidth values on the FER. 

\subsubsection{LOS Doppler Shift Variation}
\begin{figure}[!h]    
	\centering
	\includegraphics[width=0.95\columnwidth]{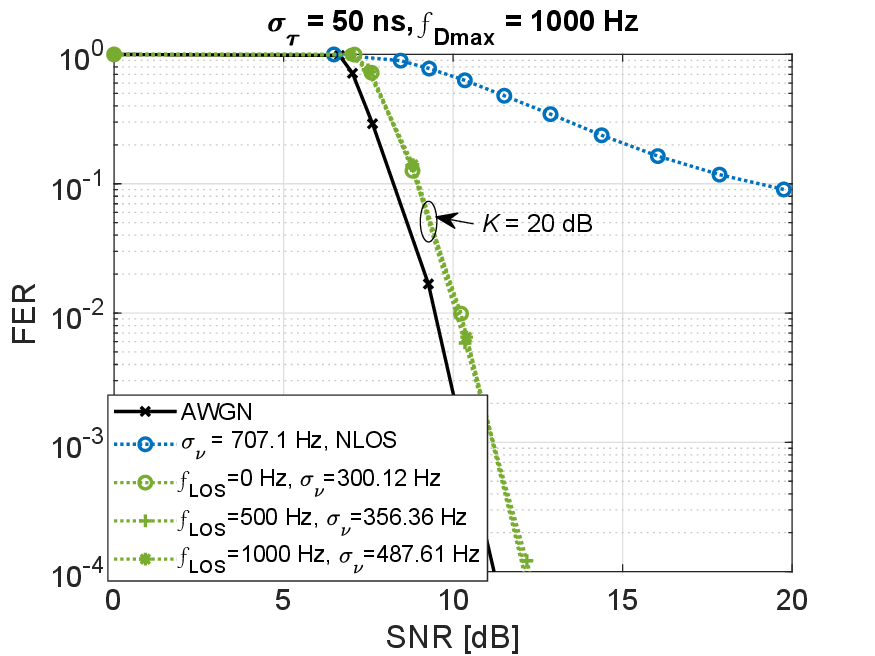}
	\caption{The FER vs. SNR measurement results for a Doppler bandwidth of $1000\,\text{Hz}$ and RMS delay spread of $50\,\text{ns}$ with $f_{\text{LOS}}=[0,\frac{f_{\text{max}}}{2},f_{\text{max}}]$. All results are compared with the FER for a pure Rayleigh fading (NLOS).}
	\label{fig:LOS1}
\end{figure}

We analytically investigated the impact of the LOS Doppler shift frequency on the RMS Doppler spread in Section \ref{subsec:TDLModel} (cf. Fig.~\ref{fig:K20FDmax2kHz}). As we can see in Fig.~\ref{fig:K20FDmax2kHz}, a higher Doppler shift of the LOS component causes a higher RMS Doppler spread. In Fig.~\ref{fig:LOS1} 
we present the FER results for different parameter configurations of the LOS Doppler shift while considering a RMS delay spread of $50\,\text{ns}$, a fixed $K$-factor of $20\,\text{dB}$ and a fixed Doppler bandwidth of $1000\,\text{Hz}$. The results show that the Doppler shift of the LOS component has no impact on the FER. Hence, in all strong LOS scenarios the carrier frequency offset compensation of the modem works reliably and provides close to AWGN performance.

\subsection{Real-time Properties}
In the previous sections we lay the foundation for developing a system-level simulation methodology pursuing the goal of obtaining the FER for a specific communication system directly from simulations in real-time. Considering the non-stationary fading process, the FER changes for each stationartiy region. Hence, in order to achieve real-time we have to obtain the FER for all communication links in a shorter time period than $T_\text{stat}$. This means that all steps shown in Fig.~\ref{fig:process}~(b) have to be done for all communication links during the duration of a stationarity region. In \cite{Rainer20} we show that our GSCM is capable to simulate a specific geometry, restrict the number of MPC to $300$ and compute all channel impulse responses for less than $T_\text{stat}$. The simulation is done in MATLAB on an Intel(R) Xeon(R) Gold 6150 CPU with 2.70 GHz. However, our approach shown in this paper requires only the calculation of the propagation path parameters. This is computationally less demanding, and less time consuming than the channel impulse response calculation. On the other hand, we additionally have to estimate the condensed radio channel parameters to index the FER lookup table during the same time period $T_\text{stat}$. Therefore, we implement the approximations of the condensed radio channel parameters into our GSCM, as is shown in Sec.~\ref{subsec:Computationparameters}. The low computational time and complexity of these approximations are independent of $T_\text{stat}$.

\subsection{Validation}
\label{subsec:validation}
In order to confirm its validity and accuracy we use data obtained by a measurement campaign in an urban intersection scenario conducted using our multi-node channel sounder~\cite{Zelenbaba20a} in the inner city of Vienna. We model this intersection scenario using our GSCM. The measurements are conducted at a center frequency of $5.9\,\text{GHz}$, using a bandwidth of $150\,\text{MHz}$ (carrier spacing of $250\,\text{kHz}$), a snapshot rate of $500\,\text{$\mu$s}$ using $\lambda/2$ dipole antennas. During the measurements, the cars are driving at most $30\,\text{km/h}$. 

\begin{figure}[!h]    
	\centering
	\includegraphics[width=0.95\columnwidth]{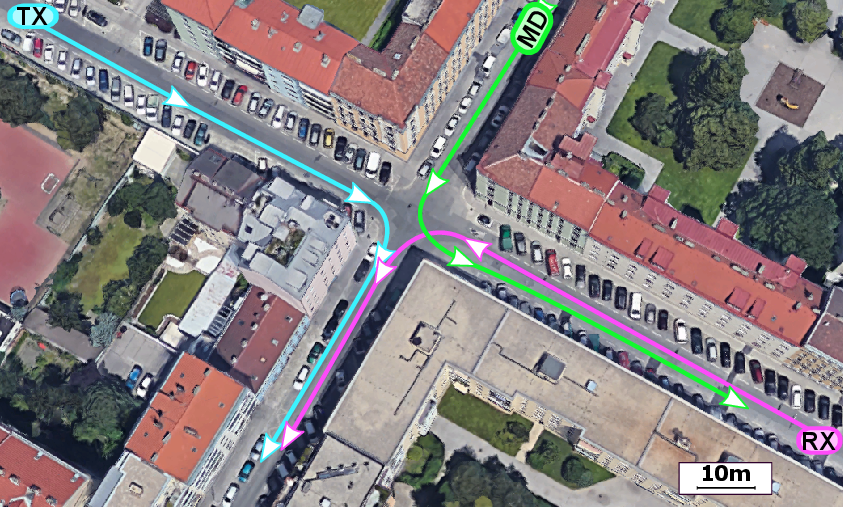}
	\caption{Measurement scenario (source: Google Maps). The trajectory of the additional car is denoted as MD.}
	\label{fig:measScenario}
\end{figure}

At the beginning, the Tx and Rx are driving toward each other establishing a LOS condition, as we can see in Fig.~\ref{fig:measScenario}. When the Tx makes a right turn at the intersection, the LOS between the Tx and the Rx is blocked by the building on the corner. The Tx and Rx establish the LOS again when the Rx turns left. 

\begin{figure}[!h]    
	\centering
	\includegraphics[width=0.95\columnwidth]{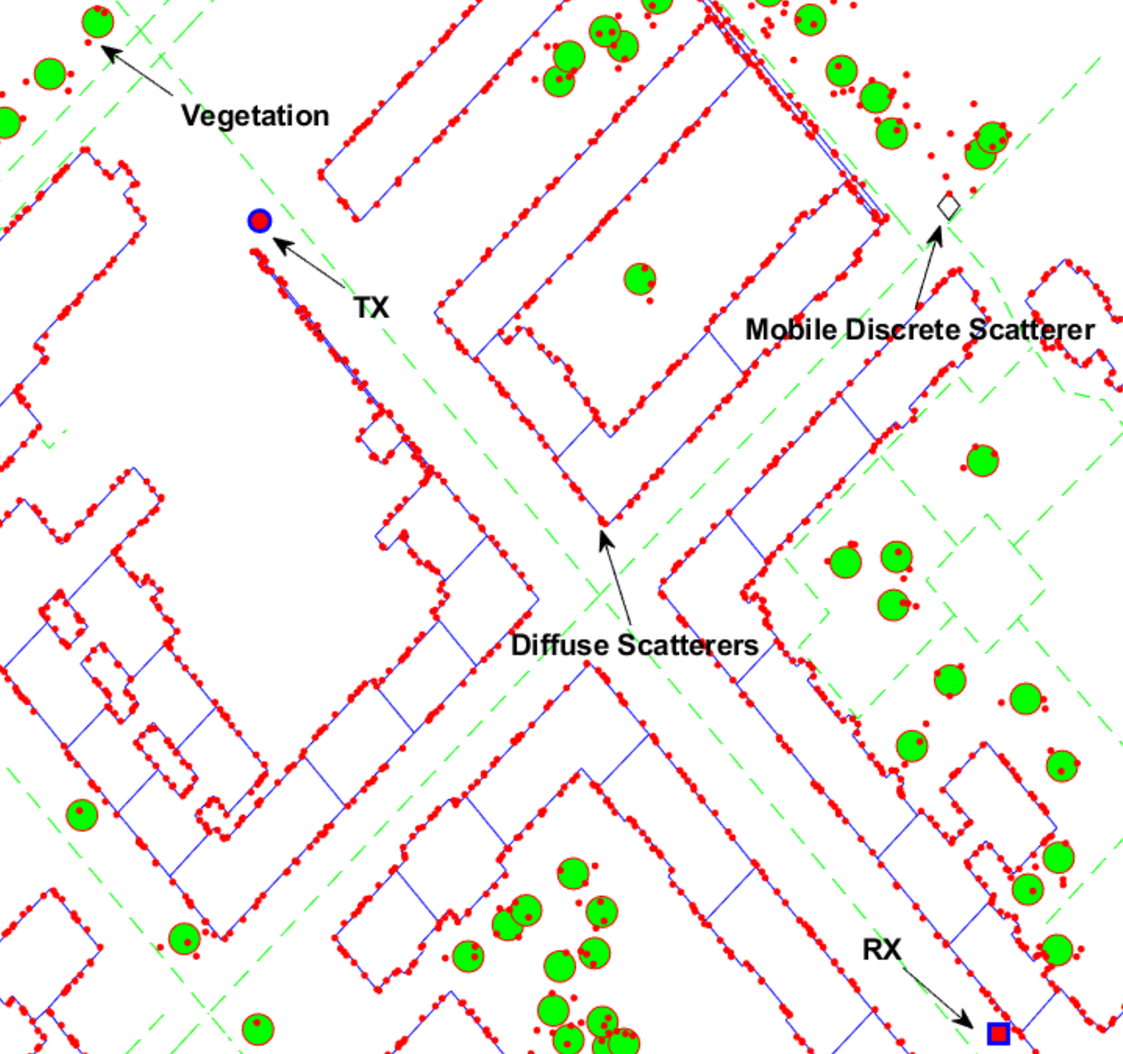}
	\caption{Automatically extracted GSCM scenario from OpenStreetMap data.}
	\label{fig:gscmScenario}
\end{figure}

Fig.~\ref{fig:gscmScenario} illustrates the reconstructed scenario with diffuse point scatterers automatically and randomly placed along buildings and vegetation. Traffic lights and signs are modeled by static discrete scatterers, whereas other moving cars that have a huge impact on the fading process in the scenario are modeled by mobile discrete scatterers. In order to extract the distribution parameters (mean, variance and coherence distance) of the large scale fading from the measurements, we follow the principles detailed in~\cite{Karedal09}. 


\subsubsection{Condensed Channel Parameter Computation from GSCM Propagation Paths}
\begin{figure}[!h]    
	\centering
	\includegraphics[width=0.95\columnwidth]{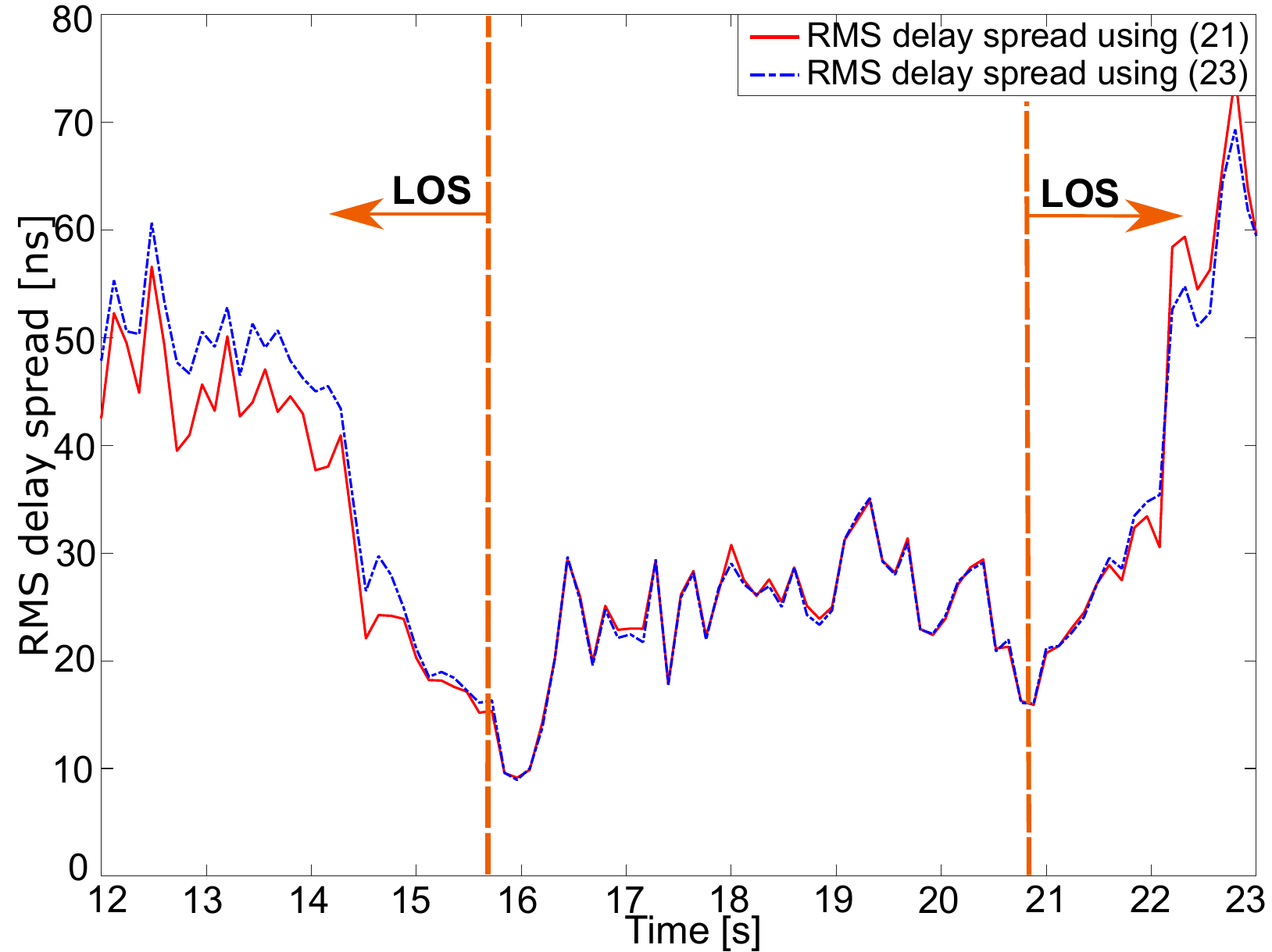}
	\caption{Comparison between the RMS delay spread obtained from PDPs defined \eqref{eq:PDP1} (solid line) and \eqref{eq:PDPSS} (dashed line) using a single simulation run.}
	\label{fig:errorinspreads2}
\end{figure}

 We evaluate the accuracy of the approximate low complexity PDP computation in \eqref{eq:PDPSS} with the ground truth from \eqref{eq:PDP1} by comparing the RMS delay spread. We use a single simulation run of the chosen scenario to investigate the introduced bias. The stationary time $T_\text{stat}$ is $120\,\text{ms}$. Fig.~\ref{fig:errorinspreads2} compares the 
 RMS delay spread over time obtained from (i) the channel impulse response \eqref{eq:PDP1} and (ii) from the propagation path parameters of the GSCM \eqref{eq:PDPSS} with low complextiy. We can see that the RMS delay spread values show the same behavior with only small deviations.
 
\begin{figure}[!h]    
	\centering
	\includegraphics[width=0.95\columnwidth]{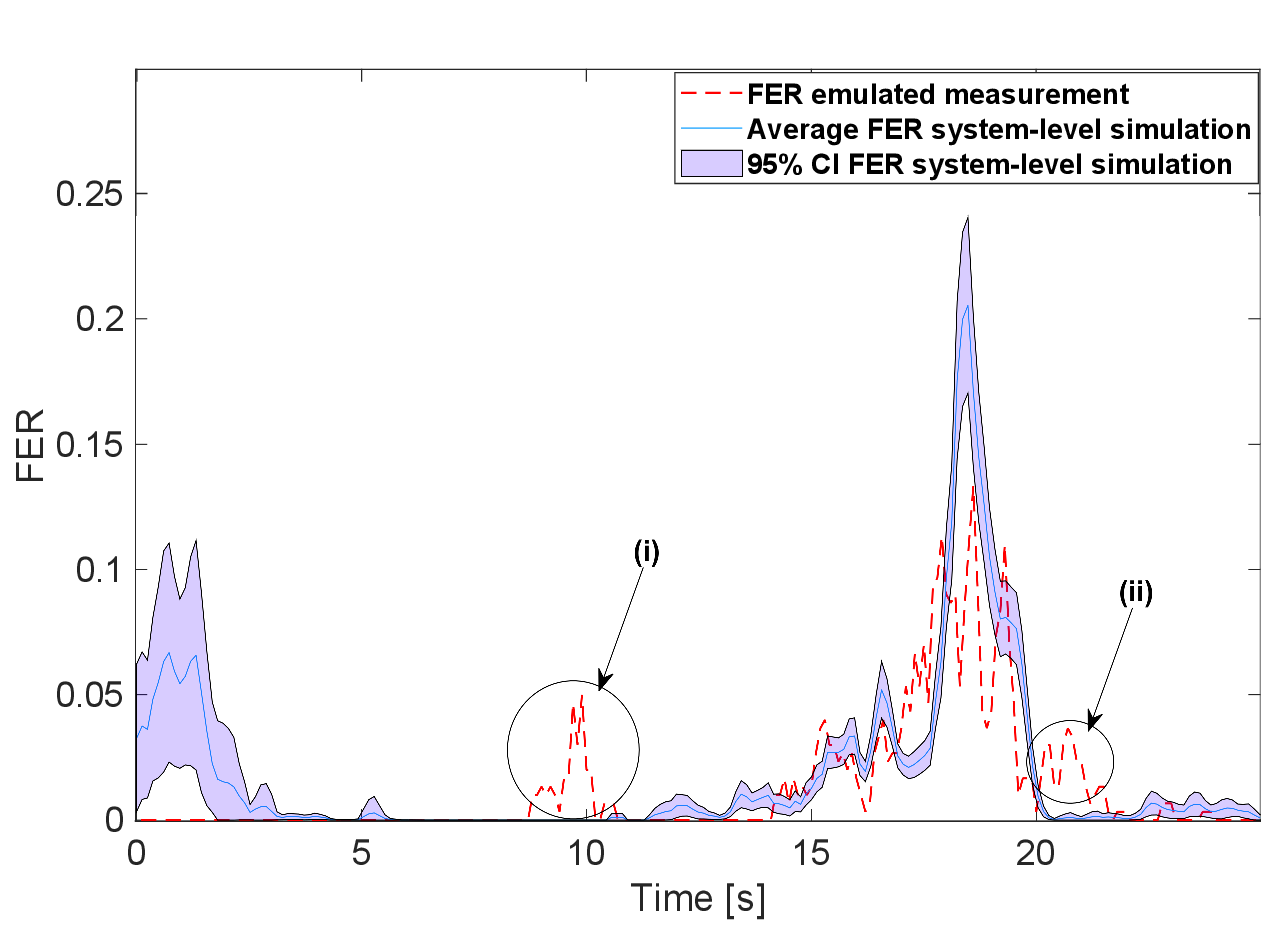}
	\caption{FERs obtained by using the FER lookup table using our GSCM compared to the FERs obtained by emulating a measurement conducted for an inner city intersection scenario.}
	\label{fig:validateFER}
\end{figure}


\subsubsection{Frame Error Rate}
Finally, we compare the FER results obtained from the system-level simulator with the FER obtained by emulating the real world measurement data, as depicted in Fig.~\ref{fig:process}~(c). We obtain the FER during system-level simulation by indexing the FER lookup table created using the methodology in Sec.~\ref{sec:methodology}. We use $100$ realizations of the GSCM and average the obtained FER. This FER is compared with the one obtained by emulating a single measurement run. The FER is calculated by averaging the FER over $100$ transmitted frames. 

Fig.~\ref{fig:validateFER} shows the average FER obtained by the system-level simulation (solid line), the corresponding $95~\%$ confidence interval (CI) (shaded area, contoured by a black solid line) and the FER we obtained by emulating the measurement data (dashed line). 

The average FER obtained by system level simulation shows an excellent match with the measured one. A difference is only visible at two time instances, with the following explanation: (i) at $10\,\text{s}$ measurement time, the attenuation is increased due to balconies which we did not model in our GSCM as we import the data directly from OpenStreetMap; (ii) at $20\,\text{s}$ measurement time parked cars cause a partial obstruction depending on their orientation and size. At all the other positions we achieve a good match.

\section{Conclusion}
\label{sec:conlcusion}
In this paper we presented a real-time system-level simulation for vehicular communication scenarios with multiple vehicles. We use a GSCM to enable a simplified but still accurate representation of the non-stationary vehicular fading process. 

The propagation path parameters of the GSCM are used to compute directly the time-variant condensed radio channel parameters per stationarity region of a communication link during run-time with low computational complexity. Further, in order to obtain real-time properties, we introduced a FER lookup table. 

A HiL test setup is used to obtain the content of the FER lookup table for different condensed radio channel parameters (path loss, RMS delay spread, Doppler bandwidth, $K$-factor and LOS Doppler shift). We measured the FER using specific IEEE 802.11p hardware modems and the AIT real-time channel emulator. We used a stochastic channel model with an exponentially decaying PDP and Clarke's Doppler spectrum with an extra LOS component. We assumed that the channel statistics of each stationarity region can be approximated by the statistics of this stochastic channel model. 

Our real-time system-level simulation is successfully validated with real world measurements data from a street crossing scenario. We obtained the FER by emulating the real word measurement data and then we compared it with the FER obtained with our system-level simulation, averaged over multiple realizations of the GSCM for that scenario. Demonstrating a good match, we showed that our approach reduces the computational complexity enabling real-time system-level simulation with good accuracy for the first time.

\section*{Acknowledgment}
This work is funded by the Austrian Research Promotion Agency (FFG) and the Austrian Ministry for Transport, Innovation and Technology (bmvit) within the project REALISM (864188) of the funding program transnational projects. 

\appendices
\section{Calculation of RMS Delay Spread for an Exponential PDP}
\label{app:RMSDs}
In order to obtain the RMSDs of the exponential PDP we use the expression from \eqref{delayspread}, where we put \eqref{pdpdiscrete} instead of $P_{\tau}$ so that we get
\begin{equation}
\sigma_{\tau} =  \sqrt{\frac {\sum_{k=0}^{N_\text{taps}-1}  (k\Delta t)^2\me^{-k\Delta t/\tau_0}}{\sum_{k=0}^{N_\text{taps}-1}\me^{-k\Delta t/\tau_0}} -(\overset{-}{\tau}})^2 
\label{eq:expdelayspread}
\end{equation}
where $\overset{-}{\tau}$ represents the mean delay and it is calculated using the expression from \eqref{meandelay}
\begin{align} \overset{-}{\tau} & = \frac{\sum_{k = 0}^{N_\text{taps}-1} k\Delta t \me^{-k\Delta t/\tau_0}}{\sum_{k=0}^{N_\text{taps}-1}\me^{-k\Delta t/\tau_0}} \nonumber \\ & =
\Delta t \frac{q^{N_\text{taps}+1}(N_\text{taps}-1)-K{_\text{taps}}q^{N_\text{taps}}+q}{(1-q)(1-q^{N_\text{taps}})}
\label{eq:expmeandelay} 
\end{align}
with $q=\me^{-\Delta t/\tau_0}$. Finally, after inserting \eqref{eq:expmeandelay} in \eqref{eq:expdelayspread} we get the following expression
\begin{equation}
     	\sigma_{\tau} =  \frac{\Delta t}{(1-q)(1-q^{N_\text{taps}})}\sqrt{A},
 	\label{eq:expdelayspreadfinal}
\end{equation}
where
\begin{align}
    A =&-2(N_\text{taps}-1)^2q^{2N_\text{taps}+2}+q^{2N_\text{taps}+1}-N_\text{taps}^2q^{N_\text{taps}+2} \nonumber \\ & +2(N_\text{taps}^2-1)q^{N_\text{taps}+1}-N_\text{taps}^2q^{N_\text{taps}}+q \nonumber
\end{align}

\section{Derivation of the Low-Complexity Radio channel parameters }
\subsection{Derivation of the PDP approximation}
\label{app:PDPapprox}

Starting from a general non-stationary doubly selective fading process 
\begin{equation}
h_\text{ph}(t,\tau) = \sum_{l=1}^L\eta_l(t)\delta(\tau-\tau_l(t)) \nonumber,
\label{phCIR_app}
\end{equation}
after sampling \eqref{discereteCIR} we can express the short-term PDP as
\begin{align}
\label{eq:PDP1_app}
&P_{\tau,r}[n] = \nonumber \\ 
& = \frac{1}{M}\sum_{m=0}^{M-1} \left|\sum_{l=1}^{L} \eta_{l}[m'] h_\text{RC}(nT_\text{c} - \tau_l[m'])\right|^2 \nonumber .
\end{align}
For sake of simplicity we omit the stationarity region index $r$, 
\begin{align}
&P_\tau[n] = \nonumber \\ 
& = \frac{1}{M}\sum_{m=0}^{M-1} \left|\sum_{l=1}^{L} \eta_l[m] h_\text{RC}(nT_\text{c} - \tau_l[m])\right|^2 \nonumber ,
\end{align}
in the following derivations.

Assuming the delay changes linearly over time and, i.e. \cite{Hofer19},
\begin{equation}
\tau_l[m]=\tau_l[0] - \frac{f_l}{f_c}mT_\text{s}
\end{equation}
and that the amplitude of the propagation paths does not change significantly during one stationarity region $|\eta_l[m]| = |\eta_l[0]|$
\begin{align}
	&P_\tau [n]\approx \nonumber \\ &\approx\frac{1}{M}\sum_{m=0}^{M-1}\left\lvert \sum_{l=1}^L \left\lvert\eta_l[0]\right\rvert\me^{\mj2\pi(\phi_l-\nu_l m)}h_\text{RC}(nT_\text{c}-\tau_l[0])\right\rvert^2 \nonumber \\
	&=\frac{1}{M}\sum_{m=0}^{M-1}\left\lvert\sum_{l=1}^L \underbrace{\left\lvert\eta_l[0]\right\rvert h_\text{RC}(nT_\text{c}-\tau_l[0])\me^{\mj2\pi\phi_l}}_{a_{l}[n]}\me^{-\mj2\pi\nu_l m}\right\rvert^2\label{eq:aldef}\\
	&=\frac{1}{M}\sum_{m=0}^{M-1}\left\lvert\sum_{l=1}^L a_{l}[n]\me^{-\mj2\pi\nu_l m}\right\rvert^2 \nonumber\\
	&=\frac{1}{M}\sum_{m=0}^{M-1}\sum_{l=1}^L\lvert a_l[n]\rvert^2 \nonumber \\ &+ \frac{1}{M}\sum_{m=0}^{M-1}\sum_{l=1}^L \sum_{k \neq l}^{L}a_l[n] a_k[n]^*\me^{-\mj2\pi\nu_l m}\me^{\mj2\pi\nu_k m} \label{PDP:last}
\end{align}

We can rewrite the mixed terms in (\ref{PDP:last}) using  $a_{l}[n]=\left\lvert\eta_l[0]\right\rvert h_\text{RC}(nT_\text{c}-\tau_l[0])\me^{\mj2\pi\phi_l}$ and $a_{k}[n]^*=\left\lvert\eta_k[0]\right\rvert h_\text{RC}(n-\tau_k[0])\me^{-\mj2\pi\phi_k}$ from (\ref{eq:aldef}) and thus we have
\begin{align}
	&\frac{1}{M}\sum_{m=0}^{M-1}\sum_{l=1}^L \sum_{k \neq l}^{L}a_l[n] a_k[n]^*\me^{-\mj2\pi\nu_l m }\me^{\mj2\pi\nu_k m } \nonumber \\
	&=\frac{1}{M}\sum_{m=0}^{M-1}\sum_{l=1}^L \sum_{k \neq l}^{L}\left\lvert\eta_l[0]\right\rvert\left\lvert\eta_k[0]\right\rvert h_\text{RC}(nT_\text{c}-\tau_l[0])  \nonumber  \\ & \cdot h^*_\text{RC}(nT_\text{c}-\tau_k[0])\me^{\mj2\pi((\nu_k-\nu_l)m +\phi_l-\phi_k)}. \nonumber
\end{align}
For easier notation we neglect $h_\text{RC}(nT_\text{c}-\tau_l[0])$ and $h^*_\text{RC}(nT_\text{c}-\tau_k[0])$ (it will be amended in the last result). For the following computation we define $\theta_{l,k} := 2\pi(\nu_k - \nu_l)$, $\gamma = \theta_{l,k} m$, and $\delta = 2\pi(\phi_l - \phi_k)$ and use the identities $\cos(\gamma)=\frac{\me^{\mj\gamma}+\me^{-\mj\gamma}}{2}$ and $\cos(\gamma + \delta)=\cos(\gamma)\cos(\delta)-\sin(\gamma)\sin(\delta)$ obtaining
\begin{align}
&\frac{1}{M} \sum_{l=1}^L\sum_{k>l}^L \left\lvert\eta_l[0]\right\rvert \left\lvert\eta_k[0]\right\rvert 2\sum_{m=0}^{M-1} \cos(\gamma+\delta). \nonumber
\end{align}
Since only $\gamma$ depends on $m$ we can write the right sum by
\begin{align}
\cos(\delta) \sum_{m=0}^{M-1} \cos(\gamma) - \sin(\delta) \sum_{m=0}^{M-1} \sin(\gamma). \nonumber
\end{align}
Writing $\cos(\gamma)=\frac{\me^{\mj\gamma}+\me^{-\mj\gamma}}{2}$ and $\sin(\gamma)=\frac{\me^{\mj\gamma}-\me^{-\mj\gamma}}{2\mj}$ in their exponential form we obtain further
\begin{align}
\label{mixed:term:intermediate}
	&\frac{1}{M} \sum_{l=1}^L\sum_{k>l}^L \left\lvert\eta_l[0]\right\rvert \left\lvert\eta_k[0]\right\rvert 2\left(\cos(\delta) \sum_{m=0}^{M-1} \frac{\me^{j\gamma}+\me^{-j\gamma}}{2}\right) \nonumber\\ &-\frac{1}{M} \sum_{l=1}^L\sum_{k>l}^L \left\lvert\eta_l[0]\right\rvert \left\lvert\eta_k[0]\right\rvert 2\left(  \sin(\delta) \sum_{m=0}^{M-1} \frac{\me^{\mj\gamma}-\me^{-\mj\gamma}}{2\mj}\right).
\end{align}
Writing back $\gamma$ as $\theta_{l,k}m$ and using the formulas for the geometric series 
\begin{align}
\label{eq:exppos}
&\sum_{m=0}^{M-1} \me^{\mj\theta_{l,k}m} = \frac{1-\me^{\mj\theta_{l,k}M}}{1-\me^{\mj\theta_{l,k}}} \nonumber \\ &= \frac{1-\left(\cos(\theta_{l,k}M) + \mj\sin(\theta_{l,k}M)\right)}{1-(\cos(\theta_{l,k}) + \mj\sin(\theta_{l,k}))}
\end{align}
\begin{align}
\label{eq:expneg}
&\sum_{m=0}^{M-1} \me^{-\mj\theta_{l,k}m} = \frac{1-\me^{-\mj\theta_{l,k}M}}{1-\me^{-\mj\theta_{l,k}}} \nonumber \\ &= \frac{1-(\cos(\theta_{l,k}M) - \mj\sin(\theta_{l,k}M))}{1-(\cos(\theta_{l,k}) - \mj\sin(\theta_{l,k}))}
\end{align}
we obtain
\begin{align}
&\frac{1}{M} \sum_{l=1}^L\sum_{k>l}^L \left\lvert\eta_l[0]\right\rvert \left\lvert\eta_k[0]\right\rvert  \nonumber \\ &\cdot \bigg(\cos(\delta)\left( \frac{1-\me^{\mj\theta_{l,k}M}}{1-\me^{\mj\theta_{l,k}}} + \frac{1-\me^{-\mj\theta_{l,k}M}}{1-\me^{-\mj\theta_{l,k}}}\right) \nonumber \\
&+ j \sin(\delta) \left( \frac{1-\me^{\mj\theta_{l,k}M}}{1-\me^{\mj\theta_{l,k}}} - \frac{1-\me^{-\mj\theta_{l,k}M}}{1-\me^{-\mj\theta_{l,k}}}\right)\bigg). \nonumber
\end{align}
Using (\ref{eq:exppos}) and (\ref{eq:expneg}) we further obtain by computation 
\begin{align}
\label{cos:geometric:series}
&\sum_{m=0}^{M-1} e^{\mj\theta_{l,k}m} + \sum_{m=0}^{M-1} e^{-\mj\theta_{l,k}m} = \nonumber \\ & = \frac{1-(\cos(\theta_{l,k}M) + \mj\sin(\theta_{l,k}M))}{1-(\cos(\theta_{l,k}) + \mj\sin(\theta_{l,k}))} \nonumber \\ &+ \frac{1-(\cos(\theta_{l,k}M) - \mj\sin(\theta_{l,k}M))}{1-(\cos(\theta_{l,k}) - \mj\sin(\theta_{l,k}))} \nonumber \\
&= 2\cos\left(\frac{1}{2} \theta_{l,k}(1-M) \right)\sin\left(\frac{1}{2}\theta_{l,k}\right)^{-1}\sin\left(\frac{1}{2}\theta_{l,k}M\right),
\end{align}
and
\begin{align}
\label{sin:geometric:series}
&\sum_{m=0}^{M-1} \me^{\mj\theta_{l,k}m} - \sum_{m=0}^{M-1} \me^{-\mj\theta_{l,k}m} = \nonumber \\ &= \frac{1-(\cos(\theta_{l,k}M) + \mj\sin(\theta_{l,k}M))}{1-(\cos(\theta_{l,k}) + \mj\sin(\theta_{l,k}))} \nonumber \\ &+ \frac{1-(\cos(\theta_{l,k}M) - \mj\sin(\theta_{l,k}M))}{1-(\cos(\theta_{l,k}) - \mj\sin(\theta_{l,k}))} \nonumber \\
&= -\mj 2\sin\left(\frac{\theta_{l,k}}{2}(1-M) \right)\sin\left(\frac{\theta_{l,k}}{2}\right)^{-1}\sin\left(\frac{\theta_{l,k}}{2}M\right). 
\end{align}
Plugging (\ref{cos:geometric:series}) and (\ref{sin:geometric:series}) into (\ref{mixed:term:intermediate}) and applying the identity $\cos(\gamma - \delta)=\cos(\gamma)\cos(\delta)+\sin(\gamma)\sin(\delta)$ we have
\begin{align}
	&\frac{1}{M} \sum_{l=1}^L\sum_{k>l}^L \left\lvert\eta_l[0]\right\rvert \left\lvert\eta_k[0]\right\rvert  \nonumber \\ &\cdot 2\sin\left(\frac{\theta_{l,k}}{2}\right)^{-1}\sin\left(\frac{\theta_{l,k}}{2}M\right) \cos\left( \Psi\right) \nonumber
\end{align}
where $\Psi =  \frac{\theta_{l,k}}{2}(1-M) - 2\pi(\phi_l - \phi_k)$. We further obtain for $\theta_{l,k}=0$ that the inner part is equal to $2\cos\left(\phi_l - \phi_k\right)$.
Writing back $\delta$ as $2\pi(\phi_l-\phi_k)$, we finally have 
\begin{align}
\label{eq:summix}
&P_\tau[n]=\sum_{l=1}^{L}\left\lvert\eta_l[0]\right\rvert^2\left\lvert h_\text{RC}(nT_\text{c}-\tau_l[0])\right\rvert^2 \nonumber \\ &+\frac{1}{M} \sum_{l=1}^L\sum_{k>l}^L \left\lvert\eta_l[0]\right\rvert \left\lvert\eta_k[0]\right\rvert h_\text{RC}[n -\tau_l[0]]h_\text{RC}[n -\tau_k[0]]  \nonumber \\
& \cdot 2
\begin{cases}
\sin\left(\frac{\theta_{l,k}}{2}\right)^{-1}\sin\left(\frac{\theta_{l,k}}{2}M\right) \cos(\Psi), \,\theta_{l,k}\neq 0 \\
M\cos\left(2\pi(\phi_l - \phi_k)\right), \, \text{otherwise} \nonumber
\end{cases}
\end{align}
\subsection{Derivation of the DSD approximation}
\label{app:DSDapprox}
In order to obtain the short-term DSD we start from \eqref{discereteCIR2}
\begin{equation}
h[m',n] \approx \sum_{l=1}^{L} |\eta_{l,r}| e^{j2 \pi (\phi_{l,r} - \nu_{l,r} m )} h_\text{RC}(nT_\text{c} - \tau_{l,r}).
\label{DSD1}
\end{equation}
Before we apply the the Fourier transform on \eqref{DSD1} with respect to $m$ we plug in our assumptions and then we compute the Doppler-variant impulse response
\begin{align}
	&s[p,n] = \sum_{m=0}^{M-1} h[m',n]\me^{-\mj 2\pi mp/M} \nonumber \\ 
	& \approx\sum_{m=0}^{M-1} \sum_{l=1}^{L} |\eta_{l,r}| \me^{\mj 2 \pi (\phi_{l,r} - \nu_{l,r} m )}h_\text{RC}(nT_\text{c}-\tau_{l,r}) \cdot \me^{-\mj 2\pi mp/M} 
	\label{DSD2}\\
	& = \sum_{l=1}^{L} |\eta_{l,r}| \me^{\mj 2 \pi \phi_{l,r}} h_\text{RC}(nT_\text{c}-\tau_{l,r}) \sum_{m=0}^{M-1} \me^{-\mj 2\pi \nu_{l,r} m }\me^{- \mj 2\pi mp/M} \nonumber \\
	& = \sum_{l=1}^{L} |\eta_{l,r}| \me^{\mj 2 \pi \phi_{l,r}} h_\text{RC}(nT_\text{c}-\tau_{l,r}) \cdot \nonumber \\
	& \cdot \text{sinc}\left( \left(f_{l,r} -\frac{p}{T_{\text{stat}}}\right)T_{\text{stat}}\right)
	\label{DSD3}
\end{align}
where  $p \in \left\lbrace -\frac{M}{2},...,\frac{M}{2}-1\right\rbrace $ represents the normalized Doppler shift index and $ \text{sinc}\left( x T_{\text{stat}}\right)$ arises from the fact that we filter the infinite time-variant impulse response with a rectangular filter in time with a duration of $T_{\text{stat}}$ \cite[Chapter 2.4]{Rao00}.

With \eqref{DSD3} we write the approximate DSD by defining $u_{l,r}[p]:=\left(f_{l,r} -\frac{p}{T_{\text{stat}}}\right)T_{\text{stat}}$ as
\begin{align}
&P_{\nu,r}[p] = \frac{1}{N}\sum_{n=0}^{N-1}|s_r[p,n]|^2 \nonumber \\
& = \frac{1}{N}\sum_{n=0}^{N-1}\left\lvert \sum_{l=1}^{L} |\eta_{l,r}| e^{\mj 2 \pi \phi_{l,r}} h_\text{RC}(nT_\text{c}-\tau_{l,r})\text{sinc}( u_{l,r}[p] )\right\rvert ^2.\label{finalDSD}
\end{align}

\section{Doppler Spectral Density}
\label{app:DSD:measChannel}
 
We switch over to a continuous form in time for deriving an analytic expression for the Doppler spectral density of~\eqref{eq:generatedChannlNormalized}.
We first compute the autocovariance (since our model is WSS-US under the assumption of additive white noise) of~\eqref{eq:generatedChannlNormalized} and then apply the Fourier transform over $\tau_\text{L}$ (the lag). We compute $h[0,n]h[\tau_\text{L},n]^*$ by
\begin{align}
& h[0,n]h[\tau_\text{L}, n]^* = \frac{K}{K+1} \me^{\mj 2 \pi f_{\text{LOS}}\tau_\text{L}} \delta_\text{K}(n-1)^2  \nonumber \\
& + \frac{1}{(L'-1)(K+1)} \nonumber \\
&\sum_{i=1}^{L'-1}\sum_{z=1}^{L'-1} \me^{\mj 2 \pi \rho_{1,i} \tau_\text{L}} \me^{\mj(\phi_{1,z}-\phi_{1,i})} \delta_\text{K}(n-1)^2  \nonumber \\ 
& + 2\sqrt{\frac{K}{(1+K)^2 (L'-1)}} \sum_{i=1}^{L'-1} \me^{\mj 2 \pi \rho_{1,i} \tau_\text{L}} \delta_K(n-1)^2  \nonumber \\
&  \cos(\phi_{\text{LOS}} - \phi_{1,i}) + \frac{1}{L'} \sum_{w=2}^{N_{\text{taps}}} \sum_{i=1}^{L'} \sum_{z=1}^{L'} \me^{\mj 2 \pi \rho_{w,i} \tau_\text{L}} \me^{\mj(\phi_{w,z}-\phi_{w,i})} \nonumber \\
&  \delta_\text{K}(n-w)^2, \nonumber
\end{align}
where $ \delta_\text{K}$ denotes the Kronecker delta for which in our case the identity $\delta_\text{K}(n-w)^2 = \delta_\text{K}(n-w)$ holds and using the fact that $\delta_\text{K} (n-i) \delta_\text{K}(n-j) = 0$, $i \neq j$. Since all random variables are independently and identically distributed, the joint probability distribution is trivially given by the fact $\mathbb{P}(X \leq x, Y \leq y)=\mathbb{P}(X \leq x)\mathbb{P}(Y \leq y)$, thus $\mathbb{E}[X Y] = \mathbb{E}[X] \mathbb{E}[Y]$, and such as the phases are all uniform random variables between $0$ and $2\pi$ we have $\mathbb{E}\left[\cos(\phi_{\text{LOS}} - \phi_{1,i})\right] = 0$ and $\mathbb{E}\left[\me^{\mj(\phi_{w,z}-\phi_{w,i})}\right] = 0$, $\phi_{w,i} \neq \phi_{w,z}$ (if the random variables coincide, which is only the case for $z=i$, then $\mathbb{E}\left[\me^{\mj(\phi_{w,i}-\phi_{w,i})}\right] = 1$). Thus, the expectation of the Rayleigh fading part reduces to $\mathbb{E}\left[\sum_{i=1}^{L'} \me^{\mj 2 \pi \rho_{w,i}}  \delta_K(n-w)\right]$, $1 \leq w \leq N_{\text{taps}}$. Finally, we obtain the following expression for the autocovariance:
\begin{align}
& \mathbb{E}[h[0,n]^*h[\tau_\text{L},n]] = |\alpha[1]|^2 \frac{K}{K+1} \me^{\mj 2 \pi f_{\text{LOS}}\tau_\text{L}}   \nonumber \\
& + |\alpha[1]|^2 \frac{1}{(K+1)} \frac{1}{2\pi} \int_{-\pi}^{\pi}  \me^{\mj 2 \pi f_\text{Dmax}\cos(\alpha) \tau_\text{L}} d\alpha \nonumber \\ 
& + \sum_{w=2}^{N_{\text{taps}}} |\alpha[w]|^2 \frac{1}{2\pi} \int_{-\pi}^{\pi} \me^{\mj 2 \pi f_\text{Dmax}\cos(\beta) \tau_\text{L}} d\beta. \nonumber
\end{align}
Applying the Fourier transform and changing integration order, results in
\begin{align}
& s[f] =  \mathcal{F}\{ \mathbb{E}[h[0,n]^*h[\tau_\text{L},n]]\}[f] = \nonumber \\
& =\int_{-\infty}^{\infty}\mathbb{E}[h[0,n]^*h[\tau_\text{L},n]] \me^{-j 2 \pi f \tau_\text{L}} d\tau_\text{L} \nonumber \\
& =|\alpha[1]|^2 \frac{K}{K+1} \delta_D(f_{\text{LOS}} -f)  d\tau\nonumber \\
& +|\alpha[1]|^2 \frac{1}{K+1} \frac{1}{2\pi} \int_{-\pi}^{\pi}\delta_\text{D}(f_\text{Dmax} \cos(\alpha) - f) d\alpha    \nonumber \\
&+\sum_{w=2}^{N_{\text{taps}}}  |\alpha[w]|^2 \frac{1}{2\pi}  \int_{-\pi}^{\pi} \delta_\text{D}((f_\text{Dmax} \cos(\beta) - f) d\beta  \nonumber \\
&= |\alpha[1]|^2 \nonumber \\ 
&\left( \frac{K}{1+K}\delta_D(f_{\text{LOS}} - f) + \frac{1}{1+K}\frac{1}{\pi (f_\text{Dmax}^2 - f^2)^{\frac{1}{2}}} \right)  \nonumber \\
&+\frac{1}{\pi (f_\text{Dmax}^2 - f^2)^{\frac{1}{2}}}\sum_{w=2}^{N_\text{taps}} |\alpha[w]|^2, \nonumber
\end{align}
where $\delta_\text{D}$ denotes the Dirac delta and we substitute $g(\beta) := f_\text{Dmax} \cos(\beta) - f$. We use the fact that $\delta_\text{D}((g(\beta)) = \sum_{\beta_n \text{simple zeros of $g(\beta)$}} \frac{\delta_\text{D}((\beta - \beta_n)}{|Dh(\beta)|}$, where $|Dh(\beta)| =  \sqrt{f_\text{Dmax}^2-f^2}$ (since $h(\beta) = 0$ $\iff$ $\beta=\cos^{-1}(f/f_\text{Dmax})$ and $|Dh(\beta)| = f_\text{Dmax} \sin(\beta) = f_\text{Dmax} \sqrt{1-\cos(\beta)^2}$, plugging in for $\beta$ yields the result.
\bibliography{CorrectedVersion_RealTimeVehicularWirelessSystemLevelSimulation}


\end{document}